\documentclass[11pt, a4paper]{article}
\usepackage{jheparxiv}
\usepackage[utf8]{inputenc}
\usepackage{amsmath}
\usepackage{amsfonts}
\usepackage{amssymb}
\usepackage{latexsym}
\usepackage{mathrsfs}
\usepackage{graphicx}
\usepackage{color}
\usepackage{slashed}
\usepackage{twistor}
\usepackage[all]{xy}

\newcommand{\sa}{\mathsf{a}}

\newcommand{\vepsilon}{\varepsilon}
\newcommand{\sA}{\mathsf{A}}
\newcommand{\sK}{\mathsf{K}}
\newcommand{\tepsilon}{\tilde{\epsilon}}
\newcommand{\tvepsilon}{{\tilde\varepsilon}}

\renewcommand{\d}{\mathrm{d}}

\subheader{\hfill \begin{tabular}{r} \texttt{IMPERIAL-TP-TA-2017-02}
 \\ \texttt{KITP-NSF-ITP-17-083} \end{tabular}}

\title{Scattering on plane waves and the double copy}

\author[a,c]{Tim Adamo,}
\author[b,c]{Eduardo Casali,}
\author[b,c]{Lionel Mason}
\author[b]{\& Stefan Nekovar}

\affiliation[a]{Theoretical Physics Group, Blackett Laboratory \\
        Imperial College London, SW7 2AZ, United Kingdom}

\affiliation[b]{The Mathematical Institute \\
        University of Oxford, Woodstock Road, OX2 6GG, United Kingdom}

\affiliation[c]{Kavli Institute for Theoretical Physics \\
        University of California, Santa Barbara, CA 93106, USA}

\emailAdd{t.adamo@imperial.ac.uk}
\emailAdd{[lmason,casali,nekovar]@maths.ox.ac.uk}

\abstract{Perturbatively around flat space, the scattering amplitudes of gravity are related to those of Yang-Mills by colour-kinematic duality, under which gravitational amplitudes are obtained as the `double copy' of the corresponding gauge theory amplitudes. We consider the question of how to extend this relationship to curved scattering backgrounds, focusing on certain  `sandwich' plane waves. We calculate the 3-point amplitudes on these backgrounds and find that a notion of double copy remains in the presence of background curvature: graviton amplitudes on a gravitational plane wave are the double copy of gluon amplitudes on a gauge field plane wave. This is non-trivial in that it requires a non-local replacement rule for the background fields and the momenta and polarization vectors of the fields scattering on the backgrounds. It must also account for new `tail' terms arising from scattering off the background. These encode a memory effect in the scattering amplitudes, which naturally double copies as well.} 

\begin{document}

\maketitle



\section{Introduction}

String theory methods have had a remarkable impact on the calculation of field theory scattering amplitudes.  In particular, in string theory, gravitational amplitudes are naturally related to the square of those for Yang-Mills. This leads to corresponding statements in field theory that are now well established at tree-level in the form of the KLT relations \cite{Kawai:1985xq}.  These relations have been extended to a notion of {\em colour-kinematic duality} or more simply {\em double copy}, in which gravity amplitudes can be obtained from Yang Mills by replacing the colour structures for Yang-Mills with their associated kinematic numerators in a specific class of representations~\cite{Bern:2008qj,Bern:2010ue,Bern:2010yg}. In this form, the double copy has been applied at increasingly high loop order, where it is instrumental in rendering the calculations feasible (e.g., \cite{Bern:2009kd,Bern:2012cd,Bern:2012gh,Bern:2013uka,Bern:2014sna}). These computations have demonstrated the inadequacy of standard techniques for determining the onset of UV divergences in supergravity~\cite{Bern:2015xsa,Bern:2017lpv,Bern:2017tuc}, and have even fueled speculations that four-dimensional $\cN=8$ supergravity could be perturbatively ultraviolet finite~\cite{Bern:2006kd,Bern:2011qn}. 
 
The double copy is a precise conjecture about how, in a specific class of representations, momentum space formulae for gravity scattering amplitudes are related to those of gauge theory.  Suppose there exist representations for which the kinematic numerators of a gauge theory scattering amplitude (expressed as a sum over cubic Feynman graphs) obey the same Jacobi-like relations as the colour factors of the amplitude. If such a set of numerators can be found, then the corresponding gravity amplitude is given by simply replacing the colour factors in the gauge theory amplitude by another copy of the kinematic factors in this gauge. At tree-level, the double copy conjecture has been proven in a number of different ways~\cite{BjerrumBohr:2009rd,Stieberger:2009hq,BjerrumBohr:2010zs,Feng:2010my,Tye:2010dd}, and is equivalent to the KLT relations~\cite{Kawai:1985xq} between open and closed string amplitudes in the low-energy limit. While there is currently no general proof at higher loop orders in perturbation theory, a growing body of evidence suggests that the double copy also holds at loop level, at the time of writing to 5 loops. The success of the double copy prescription has led to an oft-repeated slogan in the amplitudes community: Gravity = $(\mbox{Gauge Theory})^2$.

Yet, despite this array of evidence, the geometric and fully non-linear origins of the double copy remain mysterious. Most clear proofs thus far are expressed in momentum space for perturbations around a flat background.  A body of recent work has explored how to manifest the double copy at the level of classical non-linear solutions in gauge theory and gravity~\cite{Anastasiou:2013hba,Monteiro:2014cda,Luna:2015paa,Ridgway:2015fdl,Borsten:2015pla,Luna:2016due,Goldberger:2016iau,Cardoso:2016amd,Luna:2016hge,Goldberger:2017frp}. However, these studies have been restricted to algebraically special solutions (in particular those of Kerr-Schild type), and do not probe dynamics in the same way as scattering amplitudes.

In this paper, we address the question as to whether the double copy relationship between gauge theory and gravity holds for perturbation theory on curved backgrounds. To do this, we consider the simplest curved backgrounds for which there is a well-defined notion of S-matrix: \emph{sandwich plane waves}~\cite{Bondi:1958aj}. These are metric or gauge field backgrounds which are flat in the asymptotic past and future in generic directions but contain a compactly supported region of curvature. This curvature can be thought of as a burst of unidirectional radiation (gravitational or electromagnetic) which is turned on and then switched off at some finite retarded times. The possibility of scattering on a plane wave background may seem controversial in light of the fact that such space-times are not in general globally hyperbolic~\cite{Penrose:1965rx}. Nevertheless, we will see that the evolution of massless fields is unitary without leakage, so the S-matrix does indeed make sense.


The relationship Gravity =(Yang Mills$)^2$ is already nicely manifest in the underlying gravitational and electromagnetic plane waves, written in Brinkmann coordinates. With coordinates $X^{\mu}=(u,v,x^a)$, $a=1,\ldots , d-2$, the Brinkmann form of the metric is Kerr-Schild, given by 
$$
\d s^2=\d s^2_{\rm flat}  -H_{ab}(u)\,x^a\,x^b\, \d u^2\, , \qquad\mbox{ where } \qquad \d s^2_{\rm flat}=2 \d u\, \d v  - \delta_{ab}\,\d x^a \d x^b\, ,
$$
whereas the corresponding electromagnetic potential is
$$
\sA=F(u)_a\, x^a \d u\, ,
$$
so that the metric perturbation from flat space is naturally a sum of terms of the form $A\odot A$.  Here $H_{ab}(u)$ and $F_a(u)$ are curvatures and are freely prescribable functions of $u$ subject to $H_{ab}$ being trace-free for the Einstein equations to be satisfied (this restriction disappears if a dilaton is allowed).\footnote{Note that this classical double copy differs from that for the more general Kerr-Schild pp-waves considered in~\cite{Monteiro:2014cda}. There, if the Maxwell field is $\phi k_\mu$, the metric is $\d s^{2}_{\mathrm{flat}}+\phi k_\mu k_\nu$ where $k_\mu$ is a null vector and $\phi$ a solution to the transverse wave equation. Such solutions can often be considered to be longitudinal with $\phi$ playing the role of a Coulomb-like source term that is analogous to a propagator and therefore not squared. We consider plane waves with a radiative Maxwell term, so the whole Maxwell field must be squared to obtain a gravitational field.} For a sandwich wave, $H_{ab}$ and $F_{a}$ are supported in some interval $u\in[u_1,u_2]$ so that space-time and connection are flat for $u\rightarrow\pm\infty$.  
For both types of plane wave we will see that it is possible to find complete sets of polarization states for in and out momentum eigenstates for linear massless fields of integral spins.

The flat `in' and `out' regions of sandwich plane waves allow us to define the S-matrix. We focus on the special case of 3-point amplitudes; in flat space, this is where the slogan Gravity = $(\mbox{Gauge Theory})^2$ of the double copy is literally~\cite{Kawai:1985xq}:
\begin{equation*}
\cM_{3}^{\mathrm{flat}} = \left(\cA^{\mathrm{flat}}_{3}\right)^2,
\end{equation*}
where $\cM_{3}^{\mathrm{flat}}$ and $\cA^{\mathrm{flat}}_{3}$ are the 3-point gravity and gauge-theory amplitudes in Minkowski space, stripped of overall momentum conserving delta functions and coupling constants. Hence, we expect that \emph{if} there is a notion of double copy which holds in curved backgrounds, it should be most easily found at the level of 3-point amplitudes for which propagators are not yet required.

We consider such 3-point amplitudes for scalars, gauge theory and gravity on a gravitational plane wave background, and for charged scalars and gauge theory on a Yang-Mills plane wave background in any number of space-time dimensions. In each case, the computation reduces to an integral which depends on the background field; it turns out that the \emph{integrand}\footnote{This `tree-level integrand' is the equivalent of `stripping off momentum conserving delta functions' in the flat space amplitudes.} of the resulting expression carries sufficient information to determine if there is a double copy. 

We find that the 3-point amplitudes for gluons on a plane wave gauge background and for gravitons on plane wave space-times have two parts written symbolically as
$$
\mathcal{\cA}^{\mathrm{pw}}_3=F+C\, , \qquad \mathcal{M}_3^{\mathrm{pw}}=\mathcal{F}^2 - \mathcal{C}\, .
$$
Here, $F$ is precisely the flat space-time integrand for three gluon scattering, whereas $\mathcal{F}$ is the 3-gluon integrand on the gravitational plane wave background.  Thus, there is a correction term between the square of the gluon 3-point amplitude and the graviton 3-point amplitude on a plane wave metric. The flat space $F$ can be mapped to $\mathcal{F}$ after some replacements of momenta and polarization vectors by their curved (and non-constant) counterparts. These replacements are non-local on space-time and are fixed by finding solutions to the Hamilton-Jacobi equations that allow one to bring momentum eigenstates into the interior of space-time from future or past infinity in the curved case.  That it is non-local on a curved space-time is not a surprise as the double copy is only  expressed locally  on momentum space.


The correction terms $C$ and $\mathcal{C}$ arise from the `tails' formed by the linearized free fields backscattering off the background. Scalar waves propagate cleanly on a plane wave background subject to Huygens' principle~\cite{Friedlander:1975eqa}, but spin one and spin two do not~\cite{Mason:1989}. The tails of momentum eigenstates in the past pick up terms encoding the `memory' of the field through which they have passed (i.e., the integral of the field strength in the electromagnetic case). Remarkably, we find that $C^{2}\rightarrow\mathcal{C}$ with an extension of the same replacements used to relate $F$ and $\mathcal{F}$. 

Define $\widetilde \cA_{3}=F-C$ to be the gluon 3-point integrand on a gauge background with flipped sign (or colour charge) for the background gauge field, and let $\rho$ to be the replacement maps from flat to curved kinematics and gauge to gravitational background fields.  Then our double copy can be written as
$$
\cM_3 =\rho(\cA_3\widetilde\cA_3).
$$
This is strong evidence that a notion of double copy persists more generally in the presence of background curvature.

Our formulae therefore also allow a study of the\emph{ memory effect} for plane waves on the amplitude. The key ingredient in the integrand is a vielbein whose non-trivial change from past to future exemplifies the memory effect~\cite{Braginsky:1986ia,Braginsky:1987,Ludvigsen:1989kg}, which has been studied in detail for sandwich plane waves (e.g., \cite{Zhang:2017rno,Zhang:2017geq}). For a charged field on a gauge background, it gives a momentum shift from past to future infinity proportional to the integral of the field.  On a gravitational background, the linear planes that are wave fronts of a standard momentum eigenstate in the past become  diverging quartic surfaces, \emph{Dupin cyclides}, in the future \cite{Friedlander:1975eqa}.  This memory effect will also give rise to new infrared divergences that have been studied in the case of a charged field on an electromagnetic plane wave background \cite{Dinu:2012tj,Ilderton:2012qe}.

\medskip

We review the non-linear plane wave backgrounds for both gravity and gauge theory in Section~\ref{Background}. Free fields on these backgrounds are constructed in Section~\ref{FF}, where we also confirm that (for scalars, gauge theory and gravity) the S-matrix for these states is well-defined in the sense that scattering is unitary and there is no particle creation. We close this section with a brief discussion of Huygens' principle and tails. Section~\ref{GravB} contains the calculation of 3-point amplitudes and integrands for scalars, gauge theory and gravity on the gravitational plane wave background; Section~\ref{GaugeB} contains the analogous calculations for charged scalars and Yang-Mills theory on a background plane wave gauge field. In Section~\ref{TDC}, these two calculations are mapped onto each other; this map defines the double copy for 3-point amplitudes on plane wave backgrounds. We also show how the gauge theory 3-point functions on the two backgrounds are related by a double copy map which acts only on the background. Section~\ref{Discuss} concludes. In Appendix~\ref{Impulse}, we provide explicit amplitude formulae for the special case of the impulsive plane wave background. Appendix~\ref{S-matrix-integrands} contains the operational definitions of tree-level amplitude and integrand used throughout the paper.


\section{Plane Wave Backgrounds}
\label{Background}

We begin with a brief review of plane wave backgrounds in both the gravitational and gauge theoretic contexts. More thorough treatments can be found in the literature; the focus is on those features relevant to our calculations.


\subsection{Gravitational plane waves}

Non-linear plane waves are among the oldest exact solutions to the field equations of general relativity, and have many fascinating properties (c.f., \cite{Baldwin:1926,Ehlers:1962zz,griffiths1991colliding,Stephani:2003tm,Blau:2011}). These metrics describe space-times composed of pure radiation of the gravitational field itself or a Maxwell field, propagating from past to future null infinity along a given constant null direction. Our focus will be on purely gravitational plane wave metrics, which can be interpreted as a coherent superposition of gravitons. There are two standard coordinate systems: the \emph{Einstein-Rosen}~\cite{Einstein:1937qu} and the \emph{Brinkmann}~\cite{Brinkmann:1925fr} coordinates.

In Einstein-Rosen coordinates, the metric is given by:
\be\label{ER1}
\d s^2 = 2\,\d U\,\d V - \gamma_{ij}(U)\,\d y^{i}\,\d y^{j}\,,
\ee
where the indices $i,j,\ldots=1,\ldots,d-2$ and the only non-trivial metric components, $\gamma_{ij}$, depend on $U$. These coordinates are useful because they manifest many of the symmetries of the space-time which are `hidden' in the other coordinates. The metric  \eqref{ER1} clearly has Killing vectors $\frac{\partial}{\partial V}$, $\frac{\partial}{\partial y^{i}}$, and the vectors
\be\label{Sym1}
\mathcal{X}^i = y^{i}\frac{\partial}{\partial V} + F^{ij}(U)\, \frac{\partial}{\partial y^j}\,, \quad F^{ij}(U):= \int^{U}\d s\,\gamma^{ij}(s)\,,
\ee
are also Killing. The vectors $\partial_{V}$, $\partial_{i}$ and $\mathcal{X}^{i}$ form a Heisenberg algebra,
\be\label{Sym2}
\left[\cX^{i},\, \cX^{j}\right]=0\,, \qquad \left[\frac{\partial}{\partial y^i},\,\cX^{j}\right]=\delta^{j}_{i}\frac{\partial}{\partial V}\,,
\ee
so plane wave metrics are endowed with an abelian isometry group generated by translations of the constant $U$ planes as well as this (solvable) Heisenberg symmetry. We will also see that massless field equations are most easily solved in these coordinates.

The main drawback of Einstein-Rosen coordinates is that they are essentially never global coordinates: the metric will develop coordinate singularities due to the focusing of the null geodesic congruence tangent to $\p_U$~\cite{Penrose:1965rx,Bondi:1989vm}. Furthermore, the curvature and field equations are given by somewhat complicated expressions in terms of $\gamma_{ij}$. For instance, the Ricci curvature is
\begin{equation*}
R_{UU}=-\frac{\gamma^{ij}}{2}\left(\ddot{\gamma}_{ij}+\frac{1}{2}\dot{\gamma}_{ik}\gamma^{kl}\dot{\gamma}_{lj}\right)\,,
\end{equation*}
where $\dot{f}=\partial_{U}f$ for any function $f(U)$. Thus the vacuum equations impose conditions on $\gamma_{ij}$ in the form of a second-order ODE.

The Brinkmann coordinates have the advantage that they are global, and the curvature is easily identified. In the Brinkmann chart, the metric is:
\be\label{Br1}
\d s^{2}=2\,\d u\,\d v - H(u,\mathbf{x})\,\d u^2 - \d x_{a}\,\d x^{a}\,,
\ee
with indices $a,b,\ldots=1,\ldots,d-2$. In these coordinates, the $u=\mathrm{const.}$ metric is completely flat. For pp-waves $H(u,x)$ can have general $x$-dependence, but for plane waves it is constrained to be quadratic in $x^{a}$:
\be\label{Br2}
H(u,\mathbf{x})=H_{ab}(u)\,x^{a}\,x^{b}\,.
\ee
The non-vanishing Christoffel symbols in these coordinates are:
\be\label{Christoffel}
\Gamma^{a}_{uu}=-H_{ab}(u)\,x^{b}\,, \quad \Gamma^{v}_{ua}=-H_{ab}(u)\,x^{b}\,, \quad \Gamma^{v}_{uu}=-\frac{\dot{H}(u,\mathbf{x})}{2}\,,
\ee
and the non-vanishing curvature components are
\be\label{Rcurv}
R^{a}_{\:\:ubu}=-H^{a}_{b}(u)\,,
\ee
so the vacuum equations in Brinkmann coordinates simply impose that $H_{ab}$ be trace-free: $H^{a}_{a}=0$.

\medskip

The \emph{sandwich} plane wave setup is one for which $H_{ab}(u)$ is compactly supported in $u$~\cite{Bondi:1958aj}. Without loss of generality, we assume that $H_{ab}(u)\neq0$ only for $u_{1}\leq u\leq u_{2}\leq0$; for $u<u_1$ or $u> u_{2}$, the space-time is a flat. The flat region $u<u_1$ is referred to as the \emph{in-region}, while $u>u_2$ is the \emph{out-region}. See Figure 1 for a schematic of this setup.
  
\begin{figure}[t]
\centering
\includegraphics[scale=.6]{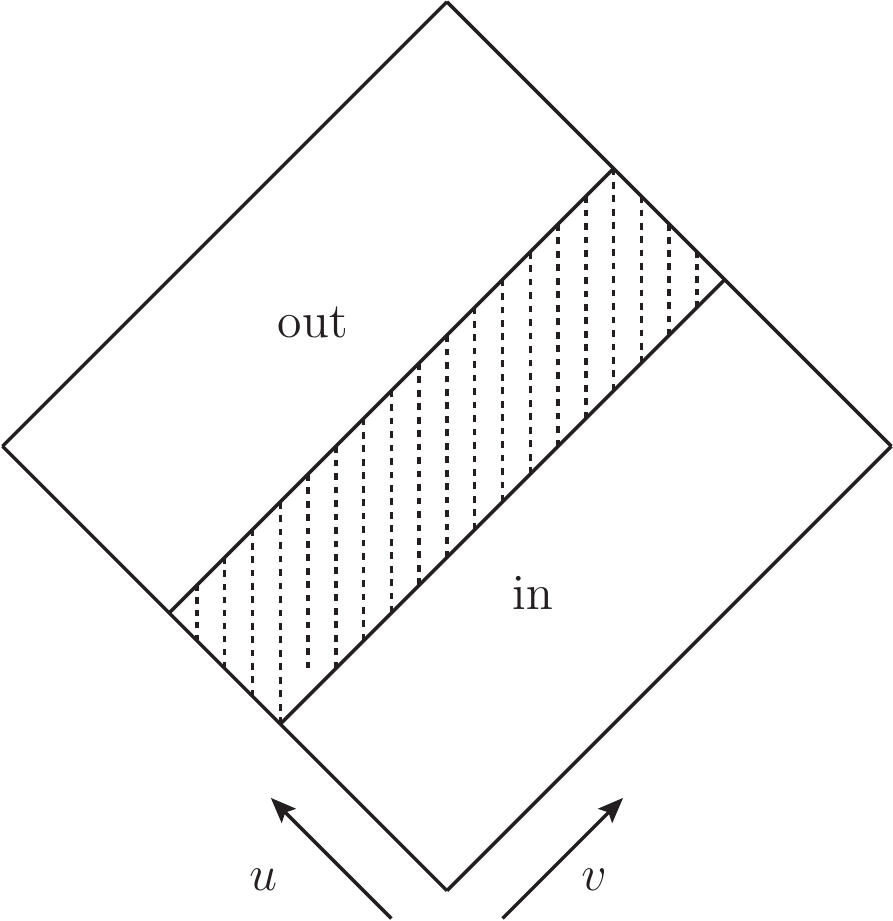}
\caption{The sandwich plane wave with $x^a$-directions suppressed. The function $H_{ab}(u)$ is non-vanishing only in the shaded region; the in- and out-regions are both flat.}
\end{figure}

Although we work mostly in Brinkmann coordinates, the relationship between the Brinkmann and Einstein-Rosen coordinate systems will be important. It can be understood in terms of the solutions to the equation:
\be\label{gde}
\ddot{e}_{a}=H_{ab}\,e^{b}\,,
\ee
for some functions $e^{a}(u)$ . Setting $e^{a}(u)=\Delta x^{a}$, \eqref{gde} is the geodesic deviation equation in Brinkmann coordinates; this follows from the fact that the connecting vectors between the geodesics,
\begin{equation*}
e^{a}\frac{\partial}{\partial x^a} - \dot{e}_{a}\,x^{a}\frac{\partial}{\partial v}\,,
\end{equation*}
are  Killing vectors. A set of $(d-2)$ Killing vectors is obtained by choosing a full $(d-2)\times (d-2)$ matrix of solutions to \eqref{gde}, $E^{a}_{i}(u)$ (and its inverse $E^{i}_{a}(u)$), subject to
\be\label{sym}
\dot{E}^{a}_{[i}\,E_{|a|\,j]}=0\,.
\ee
The Killing vectors are then:
\begin{equation*}
\mathcal{D}^{i}=E^{a\,i}\frac{\partial}{\partial x^a} - \dot{E}^{i}_{a}\,x^{a}\frac{\partial}{\partial v}\,.
\end{equation*}
The commutation relations between the $\mathcal{D}^{i}$ and the $\mathcal{X}^i$ (transformed to Brinkmann coordinates) give the Heisenberg algebra which was more manifest in Einstein-Rosen coordinates.

By comparing the line elements \eqref{ER1}, \eqref{Br1}, the diffeomorphism linking Einstein-Rosen and Brinkmann coordinates is identified as:
\begin{subequations}\label{diffeo1}
\begin{eqnarray}
U & = & u\,, \\
V & = & v +\frac{1}{2} \dot{E}^{i}_{a}\,E_{b\,i}\,x^{a}x^{b}\,, \\
y^{i} & = & E^{i}_{a}\,x^{a}\,.
\end{eqnarray}
\end{subequations}
The array $E^{a}_{i}$ and its inverse will be referred to as vielbeins since they give the $d-2$ orthonormal 1-forms $\d x^a=E^a_i\, \d y^i$ in terms of the Einstein-Rosen coordinates. They obey
\be\label{vbrel}
\ddot{E}_{a\,i}=H_{ab}\,E^{b}_{i}\,, \qquad \gamma_{ij}=E^{a}_{(i}\,E_{|a|\,j)}\,.
\ee
As part of the geometry of the Einstein-Rosen waves, the hypersurfaces $V=\mathrm{constant}$ are null and transverse to the geodesic shear-free null congruence $\p_v$ that rules the $u=\mathrm{constant}$ null hypersurfaces.  The $\p_U$ null congruence has a deformation tensor, measured in Brinkmann coordinates by
\begin{equation}
\label{shear}
\sigma_{ab}=\dot{E}^i_a\,E_{b\,i}\,,
\end{equation}
whose trace is the expansion and trace-free part is the shear.


Note that any other choice of vielbein, say $f^{a}_{i}$, is related to $E^{a}_{i}$ by
\be\label{newvb}
f^{a}_{i}=E^{a}_{j}\left( F^{jk}\,b_{ki}+c^{j}_{i}\right)\,,
\ee 
for constant matrices $b_{ij}$, $c^{i}_{j}$, and $F^{ij}(u)$ defined as:
\be\label{Fij}
F^{ij}(u):=\int^{u} \d s\,\gamma^{ij}(s) = \int^{u} \d s\,E^{a\,(i}(s)\,E^{j)}_{a}(s)\,.
\ee
In particular, given some initial value for the vielbein on the in-region of a sandwich plane wave, \eqref{newvb} encodes how the vielbein changes after passing through the curved interior to the out-region. For the sandwich wave, two natural initial values are given by requiring the vielbein to become trivial in the past or future:
\be\label{vbbc0}
\lim_{u\rightarrow\pm\infty}E^{i\,\pm}_{a}(u) = \delta_{a}^{i}\,.
\ee
Since solutions to \eqref{gde} are simply linear in flat regions, we have
\begin{equation}\label{memory}
E^{a\,-}_i(u)=b^{a\,+}_i\,u+c^{a\,+}_i \quad \mbox{as} \quad u\rightarrow +\infty\, , \quad E^{a\,+}_i(u)=b^{a\,-}_i\,u+c^{a-}_i \quad \mbox{as} \quad u\rightarrow -\infty\, .
\end{equation}
From \eqref{sym} and the conservation of the Wronskian between $E^+$ and $E^-$, it follows that 
\begin{equation}
b_{[i}^{a\,\pm}c_{j]\,a}^{\pm}=0,\qquad b^{a\,+}_i=\delta^{aj}\,\delta_{bi}\,b^{b\,-}_j
\end{equation}
and we can use a rotation of the Brinkmann coordinates to make $b$ symmetric if desired.

Note that it is essentially impossible to have $E$ invertible for all $u$ for non-trivial $b$, so the Einstein-Rosen coordinates are generically singular. This is the inevitable consequence of null geodesic focusing of the $V=\mathrm{constant}$ null hypersurfaces as emphasized by Penrose \cite{Penrose:1965rx}. Both $E^{a\,+}_{i}$ and $E^{a\,-}_{i}$ will describe the \emph{same} flat metric in the asymptotic regions but with different Einstein-Rosen forms. In particular, if the deformation tensor $\sigma_{ab}$ vanishes in one asymptotic region, it will generically be nontrivial in the other, albeit falling off as $1/u$.  
This non-trivial change in $\sigma_{ab}$  is an example of the \emph{memory effect}~\cite{Braginsky:1986ia,Braginsky:1987,Ludvigsen:1989kg}, which has been studied in detail for sandwich plane waves (e.g., \cite{Zhang:2017rno,Zhang:2017geq}).  


\subsection{Gauge theory plane waves}

An `Einstein-Rosen' plane wave in gauge theory is a gauge potential which satisfies properties similar to a plane wave metric in Einstein-Rosen coordinates. It is often used to model the electromagnetic fields of lasers (c.f., \cite{Reiss:1962,Brown:1964zzb,Ilderton:2012qe}). In particular, we demand that $\sA$ -- \emph{a priori} taking values in the adjoint of some Lie algebra $\mathfrak{g}$ -- manifests the symmetries generated by $\frac{\partial}{\partial v}$ and $\frac{\partial}{\partial x^{a}}$. The most general connection satisfying these conditions has the form:
\be\label{gER1}
\sA=\sA_{0}(u)\,\d v + \sA_{a}(u)\,\d x^{a}\,,
\ee
where we write the potential in the coordinates
\be\label{Mink}
\d s^2 =2\,\d u\,\d v - \d x_{a}\,\d x^{a}\,,
\ee
of Minkowski space. 

We want \eqref{gER1} to be preserved under the same Heisenberg symmetry algebra \eqref{Sym2} that generated the isometries of the plane wave metrics in Einstein-Rosen coordinates. This requires there to be a vector field
\be\label{gSym1}
\cX_{\varphi}^{a}=x^{a}\,\frac{\partial}{\partial v} + u\,\frac{\partial}{\partial x_{a}} +\varphi^{a}\,,
\ee
with $\varphi^{a}$ a Lie algebra-valued function for which
\be\label{gSym2}
\left[\cX^{a}_{\varphi},\, \cX^{b}_{\varphi}\right]=0\,, \qquad \left[\frac{\partial}{\partial x^{a}},\,\cX^{b}_{\varphi}\right]=\delta^{b}_{a}\frac{\partial}{\partial v}\,.
\ee
These conditions imply that $\varphi^{a}=\varphi^{a}(u)$ and $[\varphi^{a},\varphi^{b}]=0$. Furthermore, we require that $\cX^{a}_{\varphi}$ generates a further symmetry of the gauge connection; namely, that $\D=\d +\sA$ is covariantly Lie-dragged along the $\cX^{a}_{\varphi}$. This imposes further constraints on $\sA$:
\be\label{gSymCon}
\sA_{a}=-\dot{\varphi}_{a}\,, \quad  \left[\sA_{0},\,\varphi^{a}\right]=0\,, \quad \left[\sA_{a},\,\varphi^{b}\right]=\delta^{b}_{a}\,\sA_{0}\,.
\ee 
For simplicity, we restrict our attention to the special case where $\varphi^{a}$ is valued in the Cartan subalgebra $\mathfrak{h}\subset\mathfrak{g}$. With this choice, consistency of the symmetry algebra reduces to
\be\label{gSym3}
\sA_{0}=0\,, \qquad \varphi^{a}(u)=-\int^{u}\d s\,\sA^{a}(s)\,,
\ee
and the functional form of $\cX^{a}_{\varphi}$ closely resembles that of its gravitational counterpart \eqref{Sym1}.

To summarize, our definition of an `Einstein-Rosen' plane wave gauge field (valued in the Cartan of the gauge group) results in a gauge potential of the form:
\be\label{gER2}
\sA=-\sA_{a}(u)\,\d x^{a}\,,
\ee
where an overall negative sign has been included for convenience. Just as the Brinkmann form of a plane wave metric can be obtained by the diffeomorphism \eqref{diffeo1} from Einstein-Rosen form, a gauge transformation of \eqref{gER2} gives the plane wave gauge potential in `Brinkmann' form. In particular, taking $\sA\rightarrow \sA+\d(x^{a}\sA_{a})$ gives 
\be\label{gBr1}
\sA=x^{a}\,\dot{\sA}_{a}\,\d u\,.
\ee
The fact that $\sA$ is a linear polynomial in $x^{a}$, rather than a quadratic function as in the gravitational setting \eqref{Br2}, is a first glimpse of the double copy. It has already been noted that plane wave background geometries (for gauge theory and gravity) exhibit the double copy structure~\cite{Monteiro:2014cda}, although the distinction between linear and quadratic functions does not seem to have been noticed previously. Although we obtained \eqref{gBr1} from the Einstein-Rosen gauge by working in the Cartan subalgebra of the gauge group, general non-abelian plane waves also take this functional form~\cite{Coleman:1977ps}.

The field strength is
\be\label{gFS}
F=\dot{\sA}_{a}\,\d x^{a}\wedge\d u\,.
\ee
As for the Brinkmann metric, the gauge field \eqref{gBr1} directly encodes the field strength; \eqref{gFS} obeys the Maxwell equations, and hence the Yang-Mills equations when valued in the Cartan subalgebra of the gauge group.

The sandwich gauge field plane wave is analogous to that for gravity; the field strength  $F_a=\dot{\sA}_{a}(u)$ is taken to be compactly supported for $u_{1}\leq u\leq u_{2}\leq0$, so that it is flat in the in-region ($u<u_1$) and out-region ($u>u_2$). The memory effect here is associated with the fact that if $\sA$ is taken to vanish in the past, it will be constant and non-zero in the future
\be\label{gmem}
\sA_a|_{\mathrm{out}}-\sA_a|_{\mathrm{in}}=\int_{u_1}^{u_2} F_a\, \d u\,,
\ee
By analogy with the gravitational case, \eqref{gmem} can be viewed as encoding the \emph{electromagnetic memory effect}~\cite{Bieri:2013hqa} for plane wave gauge theory backgrounds. 


\section{Free Fields on Plane Wave Backgrounds and Inner Products}
\label{FF}

Amplitudes in flat space are functionals of free fields and are usually expressed as functions of momenta after being evaluated on momentum eigenstates. In curved space, such solutions are not so obviously available and it is here that we must use the special structure of plane waves. Friedlander showed that Huygens' principle remains valid for the scalar wave equation in plane wave space-times: there exist solutions with delta-function support on null hypersurfaces through every null direction~\cite{Friedlander:1975eqa}. These null hypersurfaces are level surfaces of solutions to the Hamilton-Jacobi equation, which provide curved space analogues of the function $k\cdot X$ for null vectors $k$ in Minkowski space. 

Such functions provide analogues of momentum eigenstates, and also lead to integral formulae for general solutions to the wave equation~\cite{Ward:1987ws}. Generalizing~\cite{Mason:1989}, we can raise the spin to obtain free fields of spin one and two with arbitrary polarizations, but Huygens' principle no longer holds and tails appear. Furthermore, a consequence of the memory effect will be that, unlike flat space-time, a momentum eigenstate in the past will not evolve into one in the future. Nevertheless,  we can show that, despite the lack of global hyperbolicity of plane waves \cite{Penrose:1965rx}, the scattering problem is well-defined on a plane wave background, featuring unitary evolution without leakage or particle creation. 


\subsection{Scalar wave equation}

The plane progressing waves of Friedlander are obtained from solutions to the Hamilton-Jacobi equation for null geodesics 
$$
g^{\mu\nu}(\p_\mu\phi)(\p_\nu \phi)=0\, ,
$$
such that arbitrary functions of $\phi$ satisfy the wave equation (when multiplied by a fixed pre-factor). Solutions are most easily obtained in Einstein-Rosen coordinates where they can be separated using the explicit symmetries leading to 
$$
\phi_k=k_0\, v+ k_i\,y^i +\frac{k_ik_jF^{ij}(U)}{2\,k_0}\, , 
$$
where $(k_0,k_i)$ are constants and $F^{ij}=\int \gamma^{ij}(s)\d s$ as in \eqref{Fij}. The wave equation in Einstein-Rosen coordinates is
\be\label{sweqER}
\frac{1}{\sqrt{-|g|}}\partial_{\mu}\left(\sqrt{-|g|}\,g^{\mu\nu}\,\partial_{\nu}\,\Phi\right)=\left(2\partial_{U}\,\partial_{V} +(\p_U \sqrt{\gamma})\p_V - \gamma^{ij}\partial_{i}\,\partial_{j}\right)\Phi=0\,,
\ee
and it can be seen directly that this is solved by~\cite{Friedlander:1975eqa, Ward:1987ws}
\be\label{scalsol}
\Phi(X)=\Omega(U)\,\e^{\im\,\phi_{k}}\,, \qquad \Omega(U):=|\gamma^{-1}(U)|^{1/4} = |E(u)|^{-\frac{1}{2}}\,,
\ee


In Brinkmann coordinates, the wave equation is 
\be\label{sweq}
\left(2\partial_{u}\,\partial_{v}+H(u,\mathbf{x})\,\partial^{2}_{v} - \partial_{a}\,\partial^{a}\right)\Phi=0\,,
\ee
and of course this is solved by the same $\Phi$. Using \eqref{diffeo1}, it can be expressed in Brinkmann coordinates as: 
\be\label{phi}
\phi_{k}:= \frac{k_{0}}{2}\sigma_{ab}\,x^{a}x^{b}+k_{i}E^{i}_{a}\,x^{a} + k_{0}\,v + \frac{k_{i}\,k_{j}}{2\,k_{0}} F^{ij}\,,
\ee
with $F^{ij}(u)$ and $(k_0,k_i)$ as before, and $\sigma_{ab}=\dot{E}^{i}_{a}\,E_{b\,i}$ the deformation tensor defined by \eqref{shear}. The natural momentum associated with $\phi_{k}$ is:
\begin{multline}\label{momentum}
K_{\mu}\,\d X^\mu := \d\phi_{k}=\\
 k_0\,\d v
 +\left( \frac{k_0}{2}\,\dot{\sigma}_{bc}\,x^{b}x^{c}+k_{i}\dot{E}^{i}_{b}x^{b}+\frac{k_{i}k_{j}}{2k_0}\gamma^{ij}\right)\d u+(k_{i}E^{i}_{a}+k_{0}\,\sigma_{ab}x^{b})\d x^a\,.
\end{multline}
Although $K_{\mu}$ is a $(u,x^a)$-dependent generalization of the constant momentum familiar from flat space, it is nevertheless null by construction from the Hamilton-Jacobi equation. To see this explicitly, note that $\dot{\sigma}_{bc}=\dot{E}^{i}_{b}\dot{E}_{c\,i}-H_{bc}$.

The solutions $\Phi=\Omega\e^{\im\phi_k}$ clearly reduce to on-shell momentum eigenstates when the background is Minkowski space, and hence can be chosen to do so in one or other asymptotic region. We can use this to characterize in and out scattering states in terms of $\phi_{k}$: an in state $\Phi^{-}$ is one which looks like a plane wave $\e^{\im k\cdot X}$ in the in-region ($u<u_1$), while an out state $\Phi^{+}$ looks like a plane wave in the out-region ($u>u_2$). This comes down to requiring the vielbein to become trivial in the past or the future:
\be\label{vbbc}
\lim_{u\rightarrow\pm\infty}E^{a\,\pm}_{i}(u) = \delta^{a}_{i}\,.
\ee
In terms of the solution to the Hamilton-Jacobi equations, $\phi_k$, the distinction becomes:
\be\label{phibc}
\phi^{-}_{k}|_{\mathrm{in}}= k_{0}\,v+k_{i}\delta^{i}_{a}\,x^{a}+u\,\delta^{ij}\, \frac{k_{i}k_{j}}{2k_0}=\phi^{+}_{k}|_{\mathrm{out}}\,.
\ee
The positive frequency condition on these states is simply that $k_{0}\geq0$.

\medskip

Even at the level of the free theory, some interesting facts about the S-matrix on a plane wave space-time can be derived by making use of the natural inner product between two solutions to the free equation of motion. This uses complex conjugation to turn the standard symplectic form on the space of solutions of the wave equation into an inner product:
\be\label{sip1}
\left\la \Phi_1 | \Phi_{2}\right\ra = \im \int_{\Sigma} \left(\Phi_{1}\wedge * \d \bar{\Phi}_{2} - \bar{\Phi}_{2}\wedge * \d\Phi_{1}\right)\,,
\ee
where $\Sigma$ is an arbitrary Cauchy surface. Plane wave space-times do not admit a Cauchy hypersurface~\cite{Penrose:1965rx}, but one can instead choose the foliation by hypersurfaces $\Sigma_u$ of constant $u$. In this case, the inner product gives:
\be\label{sip2}
\left\la \Phi_1 | \Phi_{2}\right\ra = \im \int_{\Sigma_{u}} \d v\,\d^{d-2}x\,\left(\Phi_{1}\,\partial_{v}\bar{\Phi}_{2} - \bar{\Phi}_{2}\,\partial_{v}\Phi_{1}\right)\,,   
\ee
evaluated at some fixed $u$. 

Consider the inner product between two positive frequency in states, say $\Phi^{-}_{1}$ and $\Phi^{-}_{2}$ with constant momentum components $\{k_{0},k_i\}$ and $\{l_{0},l_i\}$ respectively. Using \eqref{sip2}, this gives
\be\label{sbog1}
\la\Phi^{-}_{1}|\Phi^{-}_{2}\ra=2\,k_{0}\,\delta(k_{0}-l_{0})\,\delta^{d-2}(k_{i}-l_{i})\,,
\ee
with all $u$-dependence dropping out. As desired, the evolution problem underlying the scattering theory is unitary, since there is no `leakage' of momentum -- at any value of $u$ -- between the two in states.

Similarly, the inner product between a positive frequency in state and a negative frequency out state (namely $\la \Phi^{+}_{1}|\bar{\Phi}^{-}_{2}\ra$) encodes the presence of `particle creation' in the plane wave background. Without loss of generality, the inner product can be evaluated at $u=0>u_{2}$, leading to:
\begin{multline}\label{sbog2}
\left\la \Phi^{+}_{1}|\bar{\Phi}^{-}_{2}\right\ra= \delta(k_{0}+l_{0})\,(k_{0}-l_{0})\,\Omega^{-}(0)\int \d^{d-2}x\,\exp\left[\im\left(\frac{l_{0}}{2}\sigma_{ab}^{-}(0)\,x^{a}x^{b}\right.\right. \\
\left.\left.+(k_{a}+l_{i}E^{-\,i}_{a}(0))x^{a}+\frac{l_i l_j}{2l_0}F^{ij}_{-}(0)\right)\right]\,.
\end{multline}
However, the assumption of positive frequency means that $k_{0}+l_{0}\geq0$, so on the support of the overall delta function this inner product vanishes:
\be\label{sbog2*}
\left\la \Phi^{+}_{1}|\bar{\Phi}^{-}_{2}\right\ra=0\,,
\ee
confirming the well-known result that there is no particle creation for scalar QFT in plane wave space-times~\cite{Gibbons:1975jb,Garriga:1990dp}. Equivalently: positive frequency in states do not develop a negative frequency part in the out-region.

The final independent inner product is between positive frequency in and out states, $\la\Phi^{+}_{1}|\Phi^{-}_{2}\ra$. This quantity encodes the amplitude for in-to-out scattering in the plane wave space-time~\cite{Garriga:1990dp}. The inner product can again be evaluated at $u=0$:
\begin{multline}\label{sbog3}
\la\Phi^{+}_{1}|\Phi^{-}_{2}\ra=2\,k_{0}\,\delta(k_{0}-l_{0})\,\e^{-\im{s}_{l}}\,\Omega^{-}(0) \\
\times \int \d^{d-2}x\,\exp\left[\im\left((k_{a}-l_{i} E^{-\,i}_{a}(0))\,x^{a} -\frac{l_0}{2}\sigma_{ab}^{-}(0)\,x^{a}x^{b}\right)\right]\,,
\end{multline}
where the (constant) phase ${s}_l$ is defined as
\begin{equation*}
{s}_{l}:=\frac{l_i\, l_j}{2l_0}F^{ij}_{-}(0)\,.
\end{equation*}
Now, by \eqref{newvb} it follows that
\be\label{vout}
E^{-}_{ia}(u)=u\,b_{ia}+c_{ia}\,, \quad \forall u>u_2\,,
\ee
where $b$, $c$ are constant, invertible $(d-2)\times(d-2)$ matrices. This leaves a Gaussian integral to do in \eqref{sbog3}, with the result:
\be\label{bog3*}
\la\Phi^{+}_{1}|\Phi^{-}_{2}\ra=2\,k_{0}\left(\frac{2\pi}{\im\, l_{0}}\right)^{\frac{d-2}{2}}\,\delta(k_{0}-l_{0})\,\frac{\e^{-\im({s}_{l}+{r}_{k,l})}}{\sqrt{|b|}}\,,
\ee
after using the fact that $\Omega^{-}(0)=\sqrt{|c^{-1}|}$ and defining another phase
\begin{equation*}
{r}_{k,l}:=-\frac{1}{2 l_{0}}(k_{a}-l_{i}c^{i}_{a})\, c^{ak} (b^{-1})^{b}_{k}\,(k_{b}-l_{j}c^{j}_{b})\,.
\end{equation*}
As expected, this matches the result in the literature~\cite{Garriga:1990dp}.


\subsection{Spin one}

The  action for free gauge fields propagating on a plane wave space-time is
\be\label{fYM1}
S^{\mathrm{free}}[A]=\frac{1}{g^{2}}\int_{M}\d u\,\d v\,\d^{d-2}x\, \tr\!\left(\nabla_{[\mu}\,A_{\nu]}\,\nabla^{\mu}\, A^{\nu}\right)\,,
\ee
where $A_{\mu}$ is the gauge field and $\nabla$ the Levi-Civita connection. We will see that on a plane wave it is consistent to simultaneously impose both a Lorenz gauge $\nabla_{\mu}A^{\mu}=0$ and a light-cone gauge $A_{v}=0$, since $\p_v$ is Killing.  With this, the linearized equations of motion for the gauge connection are
\be\label{ymeom}
g^{\rho\sigma}\nabla_{\rho}\nabla_{\sigma} A_{\mu}=0\,, \qquad \partial_{\mu} A^{\mu}=0=A_{v}\,.
\ee
These can be solved using the $d-2$ spin-raising operators
\be\label{hraise}
\mathcal{R}^{a}:=\d u\,\delta^{ab}\frac{\partial}{\partial x^{b}} +\d x^{a}\,\frac{\partial}{\partial v}\,,
\ee
where the free index labels different possible polarization states. As tensors, the $\mathcal{R}^a$ are covariantly constant. Acting on a solution to the wave equation, $\Phi$, it is easily checked that $\mathcal{R}^a\Phi$ satisfies \eqref{ymeom}, so $\mathcal{R}^a$ is naturally a spin-raising operator (this generalizes the four-dimensional approach in~\cite{Mason:1989}).  Thus with $\Phi$ the scalar wave \eqref{scalsol} we construct the free gauge field
\be\label{ymsol}
A_{\mu}\,\d X^\mu=\frac{1}{k_0}\epsilon_a\,\mathcal{R}^a\Phi = \frac{1}{k_0}\epsilon_a \mathcal{R}^a \left(\Omega\,\e^{\im\,\phi_k}\right)\,,
\ee
where $\phi_{k}$ and $\Omega$ are as before and the polarization vector $\epsilon^a$ is constant. We can also define a `curved' $\varepsilon_\mu$ so that
\be\label{polar}
A_\mu= \varepsilon_\mu\, \Phi\,, \qquad \mbox{ where }\qquad \varepsilon_\mu\, \d X^\mu =\epsilon^a  \left(\frac{k_{j}}{k_0}E^j_a+\sigma_{ab}\,x^{b}\right) \d u +\epsilon_a \,\d x^a\,.
\ee
This satisfies the free equation of motion and gauge-fixing conditions. Similarly to its flat space counterpart, the curved polarization vector obeys
\be\label{ponshell}
\vepsilon\cdot K=g^{\mu\nu} \vepsilon_{\mu} K_{\nu}=0\,,
\ee
where $K$ is as defined in \eqref{momentum}. In the flat space limit,  $A_{\mu}$ reduces to a standard linearized plane wave $\varepsilon^{\mathrm{flat}}_{\mu}\e^{\im k\cdot X}$, with the non-trivial constant components of $\varepsilon^{\mathrm{flat}}_{\mu}$ being $\epsilon_a$.

In and out states are defined in the same way as for the scalar: an in state looks like a Minkowski plane wave in the in-region, while an out state looks like a Minkowski plane wave in the out-region.

As in the scalar case, an inner product on free gauge fields is induced by the boundary term of the action~\cite{Crnkovic:1986ex}. Restricted to a constant $u$ hypersurface, this inner product is:
\be\label{gip2}
\left\la A_{1}|A_{2}\right\ra:= \im \int_{\Sigma_u}\d v\,\d^{d-2}x\,\left(A_{1}^{\mu}\,\bar{F}_{2\,v\mu}-\bar{A}^{\mu}_{2}\,F_{1\,v\mu}\right)\,,
\ee
which is easily used to compute the three cases of interest. Assuming positive frequency for all (un-conjugated) fields, one finds:
\begin{subequations}\label{gbog}
\begin{eqnarray}
\left\la A^{-}_{1} | A^{-}_{2}\right\ra &= &2\,k_{0}\,\epsilon_1 \cdot \epsilon_2\,\delta(k_{0}-l_{0})\,\delta^{d-2}(k_{i}-l_{i})\,,\\
\left\la A^{+}_{1} | \bar{A}^{-}_{2}\right\ra &= &0\,,\\
\left\la A^{+}_{1} | A^{-}_{2}\right\ra&= &2\, k_{0}\,\left(\frac{2\pi}{\im\, l_{0}}\right)^{\frac{d-2}{2}}\,\epsilon_1 \cdot \epsilon_2\,\delta(k_{0}-l_{0})\,\frac{\e^{-\im({s}_{l}+{r}_{k,l})}}{\sqrt{|b|}}\,,
\end{eqnarray}
\end{subequations}
where $\epsilon_1 \cdot\epsilon_2 = \epsilon_1^a \epsilon_2^b \delta_{ab}$ and the phases ${s}_{l}$, ${r}_{k,l}$ are the same as the scalar case. Unsurprisingly, \eqref{gbog} indicate that the evolution problem is unitary and that there is no particle creation for gauge fields propagating on the plane wave space-time.


\subsection{Spin two}

Finally, consider linearized metric fluctuations $h_{\mu\nu}$ on the plane wave background. Assuming that the background is a solution to the vacuum Einstein equations and choosing a transverse-traceless gauge for the perturbations
\be\label{gravgfgeneric}
\nabla_{\mu}h^{\mu}_{\sigma}=0=h_{\mu}^{\mu} \, ,
\ee
the linearized Einstein equation is:
\be\label{graveom}
\nabla_{\sigma}\nabla^{\sigma}h_{\mu\nu}-2 R^{\rho}_{\:\:\:\mu\nu\sigma}\,h_{\rho}^{\sigma}=0\,,
\ee
with $R^{\rho}_{\:\:\:\mu\nu\sigma}$ the background curvature tensor. For a vacuum plane wave in Brinkmann coordinates (i.e., $H_{a}^{a}=0$), the gauge for $h_{\mu\nu}$ can be further fixed by requiring the vanishing of the $v$-components $h_{v \mu}=0$. With these conditions, the linearized equation is:
\be\label{graveom2}
g^{\mu\nu}\partial_{\mu}\partial_{\nu}h_{\rho\sigma}+4\,\delta^{u}_{(\rho} \partial_{|v|} h_{\sigma)a}\,H^{a}_{b}\,x^{b}-2\,\delta^{u}_{\rho}\delta^{u}_{\sigma}\, H^{ab}\, h_{ab}=0\,,
\ee   
where all Christoffel symbols have been written out explicitly in Brinkmann coordinates.

Solutions to \eqref{graveom2} can be constructed by acting on the massless scalar twice with the spin-raising operator \eqref{hraise}. This leads to:
\be\label{grsol2}
h_{\mu\nu}\,\d X^{\mu}\,\d X^{\nu} = \frac{1}{k_0^2}\epsilon_{a}\,\mathcal{R}^{a}\left(\epsilon_{b}\,\mathcal{R}^{b}\,\Phi\right)=
\left((\varepsilon\cdot \d X)^2 - \frac{\im}{k_{0}} \epsilon_a \epsilon_b\,\sigma^{ab} \d u^2\right)\,\Phi\,,
\ee
where $\epsilon_a$ is chosen to be null with respect to $\delta^{ab}$ to ensure that the gauge condition $h_{\mu}^{\mu}=0$ is obeyed. Note in particular the `tail' term proportional to $\epsilon_a \epsilon_b \,\sigma^{ab}$: unlike in Minkowski space-time, metric perturbations on a plane wave background do not carry a polarization which is simply the `square' of a gauge field's polarization. The reason for this is that the second spin raising operator in \eqref{grsol2} acts not only on the scalar solution (which contributes a second copy of $\varepsilon_{\mu}$) but also on the first spin raising operator (or equivalently, on the first copy of $\vepsilon_{\mu}$, which -- unlike in Minkowski space -- is not a constant vector).

Thus the perturbative double copy for plane wave backgrounds involves subtleties not present in Minkowski space. For linear perturbations around flat space,  $h_{\mu\nu}\sim A_{\mu}\odot A_{\nu}$ for  momentum eigenstates, whereas in plane wave space-times we have $h_{\mu\nu}\sim A_{\mu}\odot A_{\nu} +C_{\mu\nu}$, with  correction  $C_{\mu\nu}$ given by the last term proportional to $\sigma^{ab}$ in \eqref{grsol2}. 

The boundary term in the linearized Einstein-Hilbert action induces an inner product on metric fluctuations~\cite{Crnkovic:1986ex}:
\be\label{grip}
\left\la h_{1}|h_{2}\right\ra=\im \int_{\Sigma_u} \d v\,\d^{d-2}x\,\left(h_{1}^{\mu\sigma}\,\partial_{v}\bar{h}_{2\,\mu\sigma}-\bar{h}_{2}^{\mu\sigma}\,\partial_{v}h_{1\,\mu\sigma}\right)\,.
\ee
Once again calculating the inner products between incoming and outgoing states gives:
\begin{eqnarray}\left\la h^{-}_{1} | h^{-}_{2}\right\ra& =& 2\,k_{0}\, (\epsilon_1 \cdot \epsilon_2)^2\,\delta(k_{0}-l_{0})\,\delta^{d-2}(k_{i}-l_{i})\,,\nonumber\\
\left\la h^{+}_{1} | \bar{h}^{-}_{2}\right\ra &=& 0\,,\nonumber \\
\left\la h^{+}_{1} | h^{-}_{2}\right\ra&=& 2\, k_{0}\,\left(\frac{2\pi}{\im l_{0}}\right)^{\frac{d-2}{2}}\,(\epsilon_1 \cdot \epsilon_2)^{2}\,\delta(k_{0}-l_{0})\,\frac{\e^{-\im({s}_{l}+{r}_{k,l})}}{\sqrt{|b|}}\,.\label{grbog1}
\end{eqnarray}
So despite the `correction' term in $h_{\mu\nu}$, the physical properties of unitary evolution and no particle creation are preserved.


\subsection{Charged free fields in plane wave gauge fields}

Although we assume that the background gauge potential in \eqref{gBr1} is valued in the Cartan algebra, it couples non-trivially to free fields which are charged under the gauge group. Consider a free, charged scalar: 
\be\label{fcscal1}
S^{\mathrm{free}}[\Phi]= \frac{1}{2}\int \d u\,\d v\,\d^{d-2}x\,D_{\mu}\Phi\,\overline{ D^{\mu}\Phi}\,,
\ee
where $D_{\mu}=\partial_{\mu}-\im e\sA_{\mu}$, with $\sA_{\mu}$ the background gauge field \eqref{gBr1} and $e$ the charge of $\Phi$.  In the first instance, we will take $e$ to be a standard $\U(1)$ charge, but more generally, $\sA$ takes values in the Cartan subalgebra of some gauge group, $\Phi$ in some root space, and $e$ will then be the corresponding root and $e\sA$ the corresponding contraction with $\sA$ encoding the commutator.  The free equation of motion for the charged scalar is thus
\be\label{csceom}
D_{\mu}D^{\mu}\Phi(X)=\left(2\partial_{u}\,\partial_{v} -\partial_{a}\,\partial^{a}-2\im\,x^{a}e\,\dot{\sA}_{a}\,\partial_{v}\right)\Phi(X)=0\,.
\ee
Solutions to this `charged' wave equation are given by:
\be\label{cssol}
\Phi(X)=\e^{\im\,\tilde{\phi}_{k}}\,,
\ee
where
\be\label{cssol2}
\tilde{\phi}_{k}=k_{0}\,v +(k_{a}+e\sA_{a})\,x^{a}+\frac{1}{2\,k_{0}}\,f(u)\,.
\ee
The function $f(u)$ is the analogue of the $F^{ij}(u)$ which appeared in the gravitational case:
\be\label{ffunc}
f(u):=\int^{u}\d s\,\left(k_{a}+e\sA_{a}(s)\right)\,\left(k^{a}+e\sA^{a}(s)\right)\,.
\ee
When the background gauge field is turned off, it is easy to see that these solutions become the usual momentum eigenstates of Minkowski space.

The natural momentum associated with these scalars is defined by 
\begin{align}\label{csmom1}
{\sK}_{\mu}\,\d X^\mu & :=-\im\e^{-\im\tilde \phi_k}\,D_{\mu}\,\e^{\im\tilde{\phi}_{k}}\,\d X^\mu \nonumber \\
& = k_0 \d v+ \frac{1}{2\,k_{0}}(k_{a}+e\sA_{a})(k^{a}+e\sA^{a})\d u +(k_{a}+e\sA_{a})\d x^a\,.
\end{align}
The components of $\sK_{\mu}$ are functions of $u$, but it is easy to see that this momentum is null.

The distinction between in and out states for the charged scalar is in direct analogy with the definitions on the gravitational background. An incoming state is one which looks like a Minkowski plane wave in the in-region, while an outgoing state looks like a Minkowski plane wave in the out-region. This distinction manifests itself in the boundary conditions on $\sA$:
\be\label{gfbc}
\lim_{u\rightarrow\pm\infty}\sA^{\pm}_{a}(u)=0\,.
\ee
Note that unlike the massless scalar in the gravitational background, the exponential dependence on $x^a$ for the charged scalar is at most linear in any region.

The inner product on the charged scalars is given by
\be\label{csip1}
\left\la \Phi_{1}|\Phi_2\right\ra=\im\int_{\Sigma_u}\d v\,\d^{d-2}x\,\left(\Phi_{1}\,\partial_{v}\bar{\Phi}_{2}- \bar{\Phi}_{2}\,\partial_{v}\Phi_{1}\right)\,,
\ee
and once again there are three inner products of physical interest. These are:
 \begin{eqnarray}
 \left\la \Phi^{-}_{1}|\Phi^{-}_{2}\right\ra&=&2k_{0}\,\delta(k_{0}-l_{0})\,\delta^{d-2}\!\left(k_{a}-l_{a}\right)\,,
\nonumber \\
 \left\la \Phi^{+}_{1}|\bar{\Phi}^{-}_{2}\right\ra&=&0\,,\nonumber \\
 \left\la \Phi^{+}_{1}|\Phi^{-}_{2}\right\ra&=&2k_{0}\,\delta(k_{0}-l_{0})\,\delta^{d-2}\!\left(k_{a}-l_{a}+c_{a}\right)\,\e^{\im\,\tilde{{s}}_{l}}\,,\label{csbog1}
 \end{eqnarray}
where $c_{a}$ is the inner product of $\sA^{-}_{a}(0)$ in the Cartan subalgebra with the charge of the field.  The momentum conservation then indicates the `kick' received by the field from the memory effect.  
The phase $\tilde{{s}}_{l}$ is defined by
\begin{equation*}
 \tilde{{s}}_{l}:=\frac{f_{-}(0)}{2\,l_{0}}\,. 
\end{equation*}
The equations \eqref{csbog1} indicate that the classical S-matrix associated with this charged scalar is unitary with no particle production.

\subsection{Spin one on a gauge background}

The linearized equation of motion for a gauge field $a_{\mu}$ charged under the same gauge group as the background $\sA_{\mu}$ is:
\be\label{cgfeom}
D_{\mu}\left(D^{\mu}a^{\nu}-D^{\nu}a^{\mu}\right)+a_{\mu}\left(\partial^{\mu}\sA^{\nu}-\partial^{\nu}\sA^{\mu}\right)=0\,.
\ee
Solutions to this equation are simplified by choosing a Lorenz gauge $D_{\mu}a^{\mu}=0$ along with\footnote{This is of course not possible on a general background, but is possible here because $\p_v$ is a symmetry.} $a_{v}=0$; the latter condition actually reduces the Lorenz condition to $\partial_{\mu}a^{\mu}=0$. Solutions are then found by acting on the charged scalar solution with $\mathcal{R}^a$ as before in the gravitational case. This leads to
\be\label{cgfsol}
a_{\mu}\,\d X^{\mu}=\tilde{\epsilon}_{a}\left( \d x^a + \frac{1}{k_0}(k^a+e\sA^a)\,\d u\right)\,\e^{\im\tilde{\phi}_{k}}\,,
\ee
where $\tilde{\epsilon}_{a}$ is a (constant) $(d-2)$-dimensional vector which we will take to be null. As in the gravitational case, we define a polarisation $d$-vector $\tvepsilon_\mu$ as 
\be\label{gfpol}
\tvepsilon_\mu\, \d X^\mu = \tilde{\epsilon}_{a}\,\left(\d x^a + \frac{1}{k_0}(k^a+e\sA^a)\,\d u\right). 
\ee
This polarization is on-shell in the sense that $\sK\cdot \tvepsilon=0$.

With these gauge choices, the inner product is essentially equivalent to \eqref{gip2} giving:
\begin{eqnarray}
 \left\la a^{-}_{1}|a^{-}_{2}\right\ra&=&2k_{0}\,\tilde{\epsilon}_1\cdot \tilde{\epsilon}_2\,\delta(k_{0}-l_{0})\,\delta^{d-2}\!\left(k_{a}-l_{a}\right)\,,
 \nonumber \\
 \left\la a^{+}_{1}|\bar{a}^{-}_{2}\right\ra&=&0\,,
 \nonumber \\
 \left\la a^{+}_{1}|a^{-}_{2}\right\ra&=&2\,k_{0}\,\tilde{\epsilon}_1\cdot \tilde{\epsilon}_2\,\delta(k_{0}-l_{0})\,\delta^{d-2}\!\left(k_{a}-l_{a}+c_{a}\right)\,\e^{\im\,\tilde{{s}}_{l}}\,.
\label{gfbog}
\end{eqnarray}
So we again have a unitary classical S-matrix with no particle creation, as before.


\subsection{Huygens' principle and tails}

The wave equation in flat and plane wave space-times satisfies Huygens' principle~\cite{Friedlander:1975eqa}.  In intuitive terms, the principle states that waves can propagate in all directions without  scattering off the background metric and generating a \emph{tail}. The sharp definition is that there should exist solutions to the wave equation with delta-function support along null hypersurfaces tangent to every null direction through every point. These are simply given in the above by $\Omega\,\delta(\phi_k -c)$ where $c$ is a constant. 

This principle fails for linear fields of spin one and spin two~\cite{Mason:1989}, however.  We can construct these fields by spin raising as above. At spin one, to get a field with delta function support along $\phi_k=0$, we must start by raising the spin of a solution to the scalar wave equation of the form $\Omega\,\phi_k\,\Theta(\phi_k)$ where $\Theta$ is the Heaviside step function. With this, the corresponding spin-one potential is 
$$
A =\Theta(\phi_k)\,\frac{\epsilon_a}{k_0}\,\mathcal{R}^a \left(\Omega\,\phi_k\,\Theta(\phi_k)\right)= \Omega\,\epsilon^a \left(\d x_a +\left(\frac{k_j}{k_0}\,E^j_a + \sigma_{ab}x^b\right)\d u\right)\,\Theta(\phi_k)\, ,
$$
and the field strength is
\begin{multline}
F =\d A= \delta(\phi_k)\,\Omega\,\epsilon^a \left(\d x_a +\left(\frac{k_j}{k_0}E^j_a +\sigma_{ab}x^b\right)\,\d u\right)\wedge\d\phi_k \\
+\Theta(\phi_k)\,\Omega\,\epsilon^{a}\,\left(\sigma_{ab}\,\d x^b\wedge \d u -\sigma^{b}_{b}\,\d x_{a}\wedge\d u\right)\, .
\end{multline}
We see that the field strength has developed a tail in the second line, which is not supported at $\phi_k=0$. This tail can be thought of as the consequence of the interaction between the impulsive electromagnetic field and the gravitational background. There is a similar story for the spin-two field where one starts with $\Phi= \phi_k^3\, \Theta(\phi_k)$.  

In these examples, the tail is proportional to the shear of the $\partial_{U}$ null geodesic congruence (i.e., trace-free part of $\sigma_{ab}$). So tails are generally identified by the part of the field in which the shear appears explicitly. In the free solutions constructed above, terms contributing to the tails are readily identified: $\sigma_{ab}\,x^{b} \d u$ from $\varepsilon\cdot \d X$ at spin one and two, and the spin two correction term $C=-\frac{\im}{ k_{0}}\epsilon_a \epsilon_b\sigma^{ab}\d u^2$. However, we will see that the contributions to the tail from $\varepsilon_{\mu}$ alone actually drop out of amplitude calculations. So for spin one fields on a plane wave space-time, the tail terms do not effect the amplitude -- even though they appear explicitly in the scattering states.

This observation is perhaps related to a different definition of tails for the propagation of gauge fields on a plane wave space-times, in terms of a Green's function in~\cite{Gunther:1974,Gunther:1988}. That discussion does not give tails for gauge fields but \emph{does} for graviton propagation~\cite{Harte:2013dba}, and indeed we will see that it is the extra correction term $C$ that is important for  graviton amplitudes. 

Note that this treatment of tails does not simply extend to fields propagating on the gauge theory plane wave background because we cannot simply obtain solutions from  arbitrary functions of $\tilde \Phi$ as it now has charge. So, in the gauge background case, we will simply take the tail to be those terms in a curved polarization vector that depend explicitly on the potential $\sA$. This is consistent with the fact that such potential terms encode the memory in the asymptotic regions via \eqref{gmem}, just as the deformation tensor $\sigma_{ab}$ does on a gravitational background.


\section{3-point Amplitudes on the Gravitational Background}
\label{GravB}

We now consider the 3-point amplitudes of scalars, gauge fields and gravitons on the gravitational sandwich plane wave background. In each case, this calculation is performed by evaluating the cubic part of the action on solutions to the linearized equations of motion on the background. For each theory, the amplitude formulae are presented in terms of an integral over the $u$ variable (in Brinkmann coordinates), which cannot be done explicitly for general space-times. Stripping off the integration underlying the action integral, together with the three $\Phi$s associated with the three on-shell fields, we are left with a tree-level integrand expression which is sufficient for exploring the double copy structure of the amplitudes. See appendix~\ref{S-matrix-integrands} for further discussion of the scattering amplitudes and tree-level integrand.


\subsection{Scalars}

Consider the cubic scalar theory
\be\label{scal1}
S[\Phi]=\frac{1}{2}\int_{M} \d u\,\d v\,\d^{d-2}x\,\left(g^{\mu\nu}\partial_{\mu}\Phi\,\partial_{\nu}\Phi - \frac{\lambda}{3}\Phi^{3}\right)\,,
\ee
where $g^{\mu\nu}$ is the inverse of the plane wave metric \eqref{Br1} in Brinkmann coordinates. The 3-point amplitudes of interest are given by evaluating the cubic portion of the action\footnote{A similar calculation has been done for scalar contact interactions of arbitrary valence in certain homogeneous plane wave backgrounds~\cite{Papadopoulos:2002bg}.}
\be\label{scubic}
-\frac{\lambda}{6}\int_{M}\d u\,\d v\,\d^{d-2}x\,\Phi_{1}(X)\,\Phi_{2}(X)\,\Phi_{3}(X)\,,
\ee
where $\Phi_{{r}}(X)$ are solutions to the linearized equations of motion of \eqref{scal1} for ${r}=1,2,3$. When evaluating \eqref{scubic}, there are basically two distinct configurations which need to be considered: three in states, or one out and two in states (the other configurations are easily related to these). 

The case when all three states are incoming is the easiest. This gives
\begin{multline}\label{s3p1}
-\frac{\lambda}{6}\int_{M}\d u\,\d v\,\d^{d-2}x\,\Phi^{-}_{1}(X)\,\Phi^{-}_{2}(X)\,\Phi^{-}_{3}(X)  \\
=-\frac{\lambda}{6}\, \delta^{d-1}\!\left(\sum_{{r}=1}^{3}k_{{r}}\right)\int \d u\,|E^{-}|\,(\Omega^{-})^{3}\,\exp\left(\im F^{ij}\sum_{{s}=1}^{3}\frac{k_{{s}\,i}k_{{s}\,j}}{2k_{{s}\,0}}\right) \\
=-\frac{\lambda}{6}\,\delta^{d-1}\!\left(\sum_{{r}=1}^{3}k_{r}\right)\,\int \frac{\d u}{\sqrt{|E^{-}|}}\,\exp\left(\im F^{ij}\sum_{{s}=1}^{3}\frac{k_{{s}\,i}k_{{s}\,j}}{2k_{{s}\,0}}\right)\,. 
\end{multline}
where
\begin{equation*}
 \delta^{d-1}\!\left(\sum_{{r}=1}^{3}k_{{r}}\right):= \delta\!\left(\sum_{r=1}^{3} k_{r\,0}\right)\, \delta^{d-2}\!\left(\sum_{r=1}^{3} k_{r\,i}\right)\,.
\end{equation*}
The delta functions arise from performing the integrations in $\d v$ and $\d^{d-2}x$, with $|E^{-}|$ an overall Jacobian factor appearing in the second line. Using the relationship \eqref{scalsol} between $\Omega(u)$ and $|E|$, the various $u$-dependent factors left inside the integral can be slightly simplified in passing to the third line.

The other configuration is a bit more complicated. In this case one has
\begin{multline}\label{s3p2}
-\frac{\lambda}{6}\int_{M}\d u\,\d v\,\d^{d-2}x\,\Phi^{-}_{1}(X)\,\Phi^{-}_{2}(X)\,\Phi^{+}_{3}(X)  
=-\frac{\lambda}{6}\,\delta\!\left(\sum_{{r}=1}^{3}k_{{r}\,0}\right) \int \d u\,\d^{d-2}x\, (\Omega^{-})^{2}\Omega^{+} \times\\ 
\exp\left(\im \frac{k_{3\,0}}{2}(\sigma_{ab}^{-}-\sigma_{ab}^{+}) x^{a} x^{b} \right. 
\left. +\im\,(k_{1\,i}+k_{2\,i})E^{i\,-}_{a} x^{a} + \im\, k_{3\,i}E^{i\,+}_{a} x^{a} + \sum_{{s}=1}^{3}\frac{k_{{s}\,i}k_{{s}\,j}}{2k_{{s}\,0}} F^{ij}_{{s}}\right)\,.
\end{multline}
Due to the mixed asymptotic conditions, momentum conservation in the $v$-direction no longer eliminates the quadratic $x$-dependence from the exponential, leaving a $(d-2)$-dimensional Gaussian integral. Performing this integral leaves:
\begin{multline}\label{s3p3}
-\frac{\lambda}{6\,(k_{3\,0})^{\frac{d-2}{2}}}\,\delta\!\left(\sum_{{r}=1}^{3}k_{{r}\,0}\right) \int \d u\, (\Omega^{-})^{2}\Omega^{+}\, \sqrt{\frac{(2\pi\im)^{d-2}}{|A|}} \\
\times \exp\left(-\frac{\im}{2\,k_{3\,0}}J_{a} J_{b} (A^{-1})^{ab} +\im \sum_{{s}=1}^{3}\frac{k_{{s}\,i}k_{{s}\,j}}{2k_{{s}\,0}} F^{ij}_{{s}}\right)\,,
\end{multline}
where
\begin{equation*}
A_{ab}:= \sigma_{ab}^{-}-\sigma_{ab}^{+}\,, \qquad J_{a}:=(k_{1\,i}+k_{2\,i})\,E^{i\,-}_{a} + k_{3\,i}\,E^{i\,+}_{a}\,.
\end{equation*}
Nevertheless, applying the definition of the tree-level integrand to these results (see earlier or appendix~\ref{S-matrix-integrands}), somewhat tautologically gives the extremely simple answer
\be\label{s3pi1}
\cM_{3}(\Phi^{-}_{1}, \Phi^{-}_{2}, \Phi^{\pm}_{3})=1\,,
\ee
after stripping off a power of the coupling, overall delta-functions, and `universal' $u$-dependent functions that depend on the choice of $\Phi$'s.

This is a general feature. Although the precise form of the \emph{amplitude} will vary significantly between different configurations of incoming and outgoing states -- as in \eqref{s3p1} versus \eqref{s3p3}, the \emph{integrands} will be the same. This is the closest thing to CPT symmetry in flat space-time -- interpreted here as the ability to exchange incoming and outgoing states while simultaneously conjugating polarizations and charges -- which survives on a sandwich plane wave background.


\subsection{Gauge theory}
     
The Yang-Mills action on a curved background is:
\be\label{YM1}
S[A]=\frac{1}{g^{2}}\int_{M} \tr\left(F\wedge * F\right)\,,
\ee
where $*$ is the Hodge star and $F=[D,D]$ is the curvature of the connection $D=\nabla + A$, for $\nabla$ the Levi-Civita connection. The 3-point amplitude is given by the cubic portion of the action \eqref{YM1} evaluated on linearized states of the form \eqref{ymsol}. In the Lorenz gauge of section~\ref{FF}, the 3-point amplitude reads:
\be\label{ym3p1}
g\,f^{\sa_{1}\sa_{2}\sa_3} \int \d u\,\d v\,\d^{d-2}x\left(A_{3}^{b}\,A^{\mu}_{2}\,\partial_{\mu}A_{1\,b} - A_{2}^{b}\,A_{3}^{\mu}\,\partial_{\mu}A_{1\,b} + \mathrm{cyclic}\right)\,,
\ee
where $f^{\sa_{1}\sa_{2}\sa_3}$ are the structure constants of the gauge group. As before, there are essentially two independent configurations in which this amplitude can be evaluated: three in states or two in states and one out state. 

However, some simplifications occur in the amplitude even before the asymptotic behaviour of the states has been specified. Evaluated on general linearized free fields, \eqref{ym3p1} becomes
\be\label{ym3p2}
 g\,f^{\sa_{1}\sa_{2}\sa_3} \int \d u\,\d v\,\d^{d-2}x\,\left(\vepsilon_{1}\cdot\vepsilon_{3}\,(K_{1}\cdot\vepsilon_{2}-K_{3}\cdot\vepsilon_{2}) +  \mathrm{cyclic} 
 \right)\,\prod_{{r}=1}^{3}\Omega_{{r}}\,\e^{\im\phi_{{r}}}\,,
\ee
where the $\Omega_{{r}}$ and $\phi_{{r}}$ (${r}=1,2,3$) depend on whether the state is incoming or outgoing. Since the functional form of the integrand (i.e., the portion of this expression in the parentheses) is independent of the state configuration, it suffices to identify the integrand in the simplest configuration. As in the scalar example, this will be the all incoming configuration, since there are more delta functions in this case.

Even for the three-incoming configuration, the integrand of \eqref{ym3p2} is \emph{a priori} a function of the $x^a$ through the polarizations \eqref{polar} and momenta \eqref{momentum}. However, thanks to the identities: 
\begin{eqnarray}
\label{polrels}
K_{{r}}\cdot \varepsilon_{{s}}&=&\left\{\begin{array}{c c}
                                          0 & \mathrm{if}\;\; {r}={s} \\
                                          E^{i\,a}(k_{{r}\,0}\frac{k_{{s}\,i}}{k_{s\, 0}}\epsilon_{{s}\,a}-k_{{r}\,i}\epsilon_{{s}\,a})\quad & \mathrm{otherwise}
                                         \end{array}\right. \,,
\\
\label{polrels2}
\vepsilon_{{r}}\cdot\varepsilon_{{s}}&=&\left\{\begin{array}{c c}
                \quad    \qquad  \qquad  0 \qquad\quad\qquad& \quad \mathrm{if}\;\; {r}={s} \\
-\epsilon_{r} \cdot \epsilon_{s} &
\qquad  \mathrm{otherwise}
                                         \end{array}\right.\,,
\end{eqnarray}
it follows that the integrand is actually \emph{independent} of the $x^a$. 
This allows the $\d v$ and $\d^{d-2}x$ integrals to be done as the only dependence on these variables is in the exponential:
\begin{multline}\label{ym3p3}
 g\,f^{\sa_{1}\sa_{2}\sa_3}\,\delta^{d-1}\!\left(\sum_{{r}=1}^{3}k_{r}\right)\,\int \frac{\d u}{\sqrt{|E^-|}}\,\left(\varepsilon_{1}\cdot\vepsilon_{3}\,(K_{1}\cdot\vepsilon_{2}-K_{3}\cdot\vepsilon_{2}) + \mathrm{cyclic}
\right)\,\\ \times\, \exp\left(\im F^{ij}\sum_{{s}=1}^{3}\frac{k_{{s}\,i}k_{{s}\,j}}{2k_{{s}\,0}}\right)\,.
\end{multline}
On the support of the momentum conserving delta functions, this simplifies to 
\be\label{ym3p4}
 2g\,f^{\sa_{1}\sa_{2}\sa_3}\,\delta^{d-1}\!\left(\sum_{{r}=1}^{3}k_{r}\right)\,\int \frac{\d u}{\sqrt{|E^-|}}\,\left(\vepsilon_{1}\cdot\vepsilon_{3}\,K_{1}\cdot\vepsilon_{2} +  \mathrm{cyclic}
 \right)\, \exp\left(\im F^{ij}\sum_{{s}=1}^{3}\frac{k_{{s}\,i}k_{{s}\,j}}{2k_{{s}\,0}}\right)\,.
\ee
 As we saw for the scalar, the amplitude boils down to a $u$-integration which depends on the particulars of the background plane wave geometry. The integrand, though, is easily identified as:
\be\label{ym3p5}
\boxed{\cM_{3}(A_1,A_2,A_3)=\vepsilon_{1}\cdot\vepsilon_{3}\,K_{1}\cdot\vepsilon_{2}+   \mathrm{cyclic} 
\,.}
\ee
Note that although this has the same functional form as the flat space 3-point integrand for Yang-Mills theory, it is \emph{not} equal to the flat space result. Indeed, the integrand in this case is a function of $u$, given explicitly by
\begin{equation}
\label{ym3p6}
 \cM_{3}(A_1,A_2,A_3)= -\,\epsilon_1 \cdot \epsilon_3\,E^{i}_{a}\left(\frac{k_{1\,0}}{k_{2\, 0}}\,k_{2\,i}\,\epsilon^a_{2}-k_{1\,i}\,\epsilon^a_{2}\right) +  \mathrm{cyclic}
\end{equation}
after using \eqref{polrels}--\eqref{polrels2}. Note that the tails associated with the asymptotic states do not contribute to the amplitude, as a result of the identities \eqref{polrels}--\eqref{polrels2}.

\smallskip

The other configuration -- two incoming states and one outgoing state -- is more complicated. The primary reason for this is that the $x$-dependence of the integrand does not drop out. Assuming that the scattering states are $A^{-}_{1}$, $A^{-}_{2}$ and $A_{3}^{+}$ we now have
\begin{eqnarray}
\vepsilon_{{r}}\cdot\vepsilon_{3}&=&-\epsilon_{r} \cdot \epsilon_3\,,
\nonumber\\
K_{{r}}\cdot\vepsilon_{3}&=&\epsilon_3^a\, \left(k_{{r}\,0}\frac{k_{3\,i}}{k_{3\, 0}}E^{+\,i}_{a}-k_{{r}\,i}E^{-\,i}_{a}\right)+k_{{r}\,0}\epsilon_3^a x^{b}\,(\sigma^{+}_{ab}-\sigma^{-}_{ab})\,,\nonumber \\
K_{3}\cdot\vepsilon_{{r}}&=&\epsilon_{r}^a\, \left(k_{3\,0}\frac{k_{{r}\,i}}{k_{r\,0}}E^{-\,i}_{a}-k_{3\,i}E^{+\,i}_{a}\right)+k_{3\,0}\epsilon^a_{{r}}x^{b}\,(\sigma^{-}_{ab}-\sigma^{+}_{ab})\,,
\label{mixpol}
\end{eqnarray}
for ${r}=1,2$. The integration over $\d^{d-2}x$ is now a rather involved Gaussian integral, which has the rough structure of \eqref{s3p3} plus a derivative of this result. Since the integrand is the primary object of interest here, we will only  consider \eqref{ym3p5}.


\subsection{Gravity}

The 3-point amplitude for gravitons on the plane wave background is encoded by extracting the cubic portion of the Einstein-Hilbert action,
\be\label{EH1}
S[g]=\frac{1}{\kappa^2}\int_{M} \d^{d}X\,\sqrt{-|g|}\,R\,,
\ee
perturbed around the plane wave background metric. To do this, a recent perturbative re-writing of the Einstein-Hilbert action is useful~\cite{Cheung:2016say}. For perturbations $h_{\mu\nu}$ around a fixed background geometry $g_{\mu\nu}$, this action takes the form:
\be\label{CR1}
S[h]=\frac{1}{4\,\kappa^2}\int \d^{d}X\,\sqrt{-|g|}\left[\nabla_{\mu}\sigma_{\nu\rho}\,\nabla_{\lambda}\sigma^{\kappa\rho}\left(\sigma^{\mu\lambda}\delta^{\nu}_{\kappa} - 2\,\sigma^{\nu\lambda}\delta^{\mu}_{\kappa}\right) +\sigma^{\mu\nu}\,R_{\mu\nu}\right]\,,
\ee
where the perturbations are encoded in
\begin{equation*}
\sigma_{\mu\nu}=g_{\mu\nu}+\kappa\,h_{\mu\nu}+\frac{\kappa^2}{2}h^{2}_{\mu\nu} +\cdots\,, \qquad \sigma^{\mu\nu}=g^{\mu\nu}-\kappa\,h^{\mu\nu}+\frac{\kappa^2}{2}h^{\mu\nu}-\cdots\,,
\end{equation*}
and indices are raised and lowered with the background metric (e.g., $h^{2}_{\mu\nu}=h_{\mu\rho}g^{\rho\sigma}h_{\sigma\nu}$). On the vacuum plane wave background in Brinkmann coordinates, $|g|=-1$ and $R_{\mu\nu}=0$ so expanding \eqref{CR1} to cubic order is straightforward. This leads to the cubic term:
\be\label{CR2}
\frac{\kappa}{4}\int \d u\,\d v\,\d^{d-2}x\left(h^{\mu\nu}\nabla_{\mu}h_{\rho\sigma}\nabla_{\nu}h^{\rho\sigma}-2\,h^{\rho\nu}\nabla_{\mu}h_{\rho\sigma}\nabla_{\nu}h^{\mu\sigma}\right)\,.
\ee
We have checked that this matches the cubic contribution from expanding the standard Einstein-Hilbert action around a plane wave background.

The 3-point amplitude is given by evaluating \eqref{CR2} on three of the linearized perturbations \eqref{grsol2}. With the transverse-traceless gauge conditions on $h_{\mu\nu}$, the covariant derivatives in \eqref{CR2} reduce to partial derivatives, leaving:
\be\label{gr3p1}
\frac{\kappa}{4}\int \d u\,\d v\,\d^{d-2}x\left(h_{1}^{\mu\nu}\partial_{\mu}h_{2\,\rho\sigma}\partial_{\nu}h_{3}^{\rho\sigma}-2\,h_{1}^{\rho\nu}\partial_{\mu}h_{2\,\rho\sigma}\partial_{\nu}h_{3}^{\mu\sigma}\right) + \mathrm{all}\:\:\mathrm{permutations}\,.
\ee 
A computation gives a typical term in the sum over permutations of external states to be:
\begin{multline}\label{gr3p2}
h_{1}^{\mu\nu}\partial_{\mu}h_{2\,\rho\sigma}\partial_{\nu}h_{3}^{\rho\sigma}-2\,h_{1}^{\rho\nu}\partial_{\mu}h_{2\,\rho\sigma}\partial_{\nu}h_{3}^{\mu\sigma}= \\
\Bigg(\left(2\vepsilon_{3}\cdot K_{2}\,\vepsilon_{1}\cdot K_{3}\,\vepsilon_{1}\cdot\vepsilon_{2}-\vepsilon_{1}\cdot K_{2}\, \vepsilon_{1}\cdot K_{3}\,\vepsilon_{2}\cdot\vepsilon_{3}\right)\,(\vepsilon_{2}\cdot\vepsilon_{3})
\\
-\im \,\vepsilon_{2}\cdot\vepsilon_{3}\,\sigma^{ab}\left(\frac{k_{2\,0}k_{3\,0}}{k_{1\,0}}\vepsilon_{2}\cdot\vepsilon_{3}\, \epsilon_{1\,a}\epsilon_{1\,b} -2k_{2\,0}\,\vepsilon_{1}\cdot\vepsilon_{2}\,\epsilon_{1\,b}\epsilon_{3\,a}\right)\Bigg)\,\e^{\im(\phi_{1}+\phi_{2}+\phi_{3})}\,.
\end{multline}
To proceed further, the configuration of the external states must be specified. Building on the scalar and gauge theory calculations, it is clear that the easiest configuration to treat is the one with all three states incoming.

In this configuration, identities of the form \eqref{polrels}--\eqref{polrels2} ensure that the only $x$-dependence in terms  like \eqref{gr3p2} is in the overall exponential. This allows the $\d v$ and $\d^{d-2}x$ integrations to be done explicitly, resulting in momentum conserving delta functions. On the support of these delta functions, the 3-point amplitude for incoming states reads:
\begin{multline}\label{gr3p3}
\frac{\kappa}{2}\,\delta^{d-1}\!\left(\sum_{{r}=1}^{3}k_{r}\right)\,\int \frac{\d u}{\sqrt{|E^-|}}\,\left[\left(\vepsilon_{1}\cdot\vepsilon_{3}\,K_{1}\cdot\vepsilon_{2}+ \mathrm{cyclic}
\right)^{2}\right. 
-\im\, k_{1\,0}k_{2\,0}k_{3\,0}\,\sigma^{ab}\mathcal{C}_{a}\mathcal{C}_{b}
\Big]\\
\times\, \exp\left(\im F^{ij}\sum_{{s}=1}^{3}\frac{k_{{s}\,i}k_{{s}\,j}}{2k_{{s}\,0}}\right)\,.
\end{multline}
where the quantity $\mathcal{C}_{a}$ is defined as
\be\label{curvcor}
\mathcal{C}_{a}:= \vepsilon_{2}\cdot\vepsilon_{3}\,\frac{\epsilon_{1\,a}}{k_{1\, 0}} + \vepsilon_{1}\cdot\vepsilon_{3}\, \frac{\epsilon_{2\,a}}{k_{2\, 0}}+\vepsilon_{1}\cdot\vepsilon_{2}\, \frac{\epsilon_{3\,a}}{k_{3\, 0}\,.}
\ee
The upshot is that the 3-point integrand for gravity on a plane wave space-time is given by
\begin{multline}\label{gr3p4}
\cM_{3}(h_{1},h_{2},h_{3})=\left(\vepsilon_{1}\cdot\vepsilon_{3}\,K_{1}\cdot\vepsilon_{2}+\vepsilon_{1}\cdot\vepsilon_{2}\,K_{2}\cdot\vepsilon_{3} +\vepsilon_{2}\cdot\vepsilon_{3}\,K_{3}\cdot\vepsilon_{1}\right)^{2} 
\\
-\im\, k_{1\,0}k_{2\,0}k_{3\,0}\,\sigma^{ab}\,\mathcal{C}_{a}\,\mathcal{C}_{b}\,,
\end{multline}
This structure mirrors what one might have guessed based solely on the structure of the linearized perturbations \eqref{grsol2}. 

So it seems that 3-point amplitudes on a plane wave space-time do not simply obey double copy as they do in flat space. Indeed, we find that
\be\label{dc1}
\boxed{\cM_{3}(h_{1},h_{2},h_{3})=\left(\cM_{3}(A_{1},A_{2},A_{3})\right)^{2}-\im\, k_{1\,0}k_{2\,0}k_{3\,0}\,\sigma^{ab}\,\mathcal{C}_{a}\,\mathcal{C}_{b}\,.}
\ee
Unlike the gluon amplitudes, the tails associated to graviton perturbations \emph{do} contribute to the amplitude. Note that they do so in an intrinsically geometric way: the tail contribution couples via the deformation tensor associated with the background geometry. To find the `square root' of perturbative gravity on a plane wave background, one must instead turn to Yang-Mills theory in the presence of a background plane wave gauge field.


\section{3-point Amplitudes on the Gauge Field Background}
\label{GaugeB}

The 3-point amplitudes for charged scalars and Yang-Mills theory in a plane wave background gauge field are now computed. As in the gravitational setting, these amplitudes reduce to an integral over the $u$-coordinate which depends on the particulars of the background, but the tree-level integrands are easily identified.


\subsection{Charged scalars} 

To obtain a gauge invariant cubic scalar interaction that carries charge with respect to the background gauge field, the charges of the three fields must add up to zero.  
\be\label{cscal1}
S_{\mathrm{int}}[\Phi]= \int \d u\,\d v\,\d^{d-2}x\,\left( \Phi_1\Phi_2\Phi_3\right)\,,
\ee
where $D_{\mu}\Phi_{r}=(\partial_{\mu}-\im e_{r}\sA_{\mu})\Phi_{r}$, with $\sA_{\mu}$ the background gauge field \eqref{gBr1}. The charges $e_{r}$ as roots encode the commutators.

Armed with the linearized solutions \eqref{cssol}, we can compute the 3-point amplitudes by evaluating the cubic portion of the action \eqref{cscal1}. This means that the amplitude can be reduced to a $u$-integration fairly straightforwardly in an arbitrary configuration:
\be\label{cs3p1}
\delta^{d-1}\!\left(\sum_{{r}=1}^{3}k_{{r}}\right)\,\int \d u\,\exp\left(\im\sum_{{s}=1}^{3}\frac{f_{{s}}}{2k_{{s}\,0}}\right)\,.
\ee 
Note that the translation action of the gauge field on the total momentum has cancelled because the charges must add up to zero by gauge invariance.  From this expression it is easy to read off the tree-level integrand for the 3-point scattering of charged scalars on the plane wave gauge field background:
\be\label{cs3p2}
\cA_{3}(\Phi_{1},\Phi_{2},\Phi_{3})=1\,.
\ee
This is independent of the specifics of the configuration as for  the gravitational background.


\subsection{Gauge theory}

Now consider a dynamical gauge field $a$ on the fixed plane wave background $\sA$. Although the background gauge field $\sA$ is valued in the Cartan of the gauge group, the dynamical gauge field carries arbitrary colour structure. The dynamical gauge field is governed by the action
\be\label{cgf1}
S[a]=\frac{1}{g^2}\int\tr\left(\cF\wedge*\cF-\d\sA\wedge*\d\sA\right)\,,
\ee
where $\cF$ is the curvature of $\sA+a$ and the kinetic term for the non-dynamical background field is subtracted. 

The cubic term in the action \eqref{cgf1} is
\be\label{cgf3p1}
\int \d u\,\d v\,\d^{d-2}x\,\tr\left(a_{\mu}\,a_{\nu}\left(\partial^{\mu}a^{\nu}-\partial^{\nu}a^{\mu}+[\sA^{\mu},a^{\nu}]\right)\right)\ \,.
\ee
We must choose the colour structure so as to obtain a non-trivial trace. All non-trivial examples are essentially the same and are equivalent to taking the $\SU(2)$ case with $a_3$ in the Cartan, and $a_1$, $a_2$ respectively of charge $\pm 1$ with respect to the Cartan generator. In particular the three charges add up to zero. Together with the gauge choices made in \eqref{cgfsol}, the 3-point amplitude reduces to
\be\label{cgf3p2}
g\,f^{\sa_{1}\sa_{2}\sa_3} \int \d u\,\d v\,\d^{d-2}x\,\left(a_{2}^{\mu}\,a^{\nu}_{3}\partial_{\mu}a_{1\,\nu}-a_{2}^{\mu}\,a_{3}^{\nu}\partial_{\nu}a_{1\,\mu} + \mathrm{cyclic}\right)\,.
\ee
Evaluating on the states \eqref{cgfsol} (with arbitrary asymptotics) leads to
\be\label{cgf3p3}
\im g\,f^{\sa_{1}\sa_{2}\sa_3}\,\delta^{d-1}\!\left(\sum_{{r}=1}^{3}k_{{r}}\right) \int \d u\, 
\left[\tvepsilon_{1}\cdot\tvepsilon_{3}\,({\sK}_{1}\cdot\tvepsilon_{2}-{\sK}_{3}\cdot\tvepsilon_{2})+ \mathrm{cyclic}
\right]\,\exp\left(\im\sum_{{s}=1}^{3}\frac{f_{{s}}}{2\,k_{{s}\,0}}\right)\,.
\ee
On the support of these delta functions, the result further reduces to:
\begin{equation}
\label{cgf3p4}
2\im g\,f^{\sa_{1}\sa_{2}\sa_3}\,\delta^{d-1}\!\left(\sum_{{r}=1}^{3}k_{{r}}\right) \int \d u\,  
\left[\tvepsilon_{1}\cdot\tvepsilon_{3}\,{\sK}_{1}\cdot\tvepsilon_{2}+\mathrm{cyclic}
\right]\,\exp\left(\im\sum_{{s}=1}^{3}\frac{f_{{s}}}{2\,k_{{s}\,0}}\right)\,.
\end{equation}
 Thus the integrand can be written in terms of on-shell data:
\be\label{cgf3p5}
\boxed{\cA_{3}(a_{1},a_{2},a_{3})=\tvepsilon_{1}\cdot\tvepsilon_{3}\,\sK_{1}\cdot\tvepsilon_{2}+\mathrm{cyclic}
\,,}
\ee
as expected.

This formula hides explicit dependence on the potential.
Using \eqref{csmom1} and \eqref{gfpol}, it follows that:
\begin{eqnarray}
\label{gpolrels}
\sK_{{r}}\cdot \tvepsilon_{{s}}&=&\left\{\begin{array}{c c}
                                          0 & \mathrm{if}\;\; {r}={s} \\
             \frac{ \tilde{\epsilon}_{{s}}^{a}}{k_{s 0}}(k_{{r}\,0}{k_{{s}\,a}}-{k_{s\, 0}}k_{{r}\,a}+ {k_{{r}\,0}} e_s\sA_{a}- {k_{s\, 0}}e_r\sA_{a}) \quad & \mathrm{otherwise}
                                         \end{array}\right. \,,\\
\label{gpolrels2}
\tvepsilon_{{r}}\cdot\tvepsilon_{{s}}&=&\left\{\begin{array}{c c}
                                          0 & \mathrm{if}\;\; {r}={s} \\
                                          -\tilde{\epsilon}_{{r}}\cdot\tilde{\epsilon}_{{s}} & \mathrm{otherwise}
                                         \end{array}\right.\,.
\end{eqnarray}
In particular, the background gauge field \emph{does} enter into the functional form of the integrand \eqref{cgf3p5}. The explicit form of the integrand is:
\begin{multline}\label{cgf3p6}
 \cA_{3}(a_{1},a_{2},a_{3})=-\frac{\tilde{\epsilon}_{1}\cdot \tilde{\epsilon}_{3}}{{k_{2\,0}}}\left[\left({k_{1\,0}}\,k_{2}\cdot \tilde{\epsilon}_{2}-{k_{2\,0}}\, k_{1}\cdot \tilde{\epsilon}_{2}\right) 
 +
\sA\cdot\tilde{\epsilon}_{2}\left({k_{1\,0}}e_2- {k_{2\, 0}}e_1\right)\right]
+\mathrm{cyclic}\,.
\end{multline}
Crucially, the terms linear in $\sA$ give a background-dependent correction to the flat space result analogous to the tail terms involving $\sigma_{ab}$ appearing in the gravity integrand \eqref{gr3p4}.  In both cases, they encode the memory. 



\section{The Double Copy}
\label{TDC}

Armed with explicit formulae for the 3-point integrands on both gravitational and gauge theory plane wave backgrounds, a precise statement of double copy can now be made. From \eqref{cgf3p6}, the 3-point integrand for gluons on the gauge theory plane wave background can be written compactly as:
\be\label{dc1*}
 \cA_{3}(a_{1},a_{2},a_{3})= F(\{k_{{r}\,0},k_{{r}\,a},\tilde{\epsilon}_{{r}}\}) + \,\mathsf{C}(\{k_{{r}\,0},k_{{r}\,a},\tilde{\epsilon}_{{r}}\}|\sA)\,,
\ee
where the function
\begin{equation}\label{flatamp}
F(\{k_{{r}\,0},k_{{r}\,a},\tilde{\epsilon}_{{r}}\}):=-\frac{\tilde{\epsilon}_{1}\cdot \tilde{\epsilon}_{3}}{k_{2\, 0}}\left(k_{1\,0}\,k_{2}\cdot \tilde{\epsilon}_{2}-k_{2\,0}\,k_{1}\cdot \tilde{\epsilon}_{2}\right) +\mathrm{cyclic}
\end{equation}
is the `flat' contribution to the integrand.\footnote{The spurious poles in $k_0$ are associated with our projection of the polarization vectors $\epsilon_a$ to be orthogonal to both $\p_u$ and $\p_v$.} The tail-dependent correction term is
\begin{equation}\label{gcorr1}
\mathsf{C}(\{k_{{r}\,0},k_{{r}\,a},\tilde{\epsilon}_{{r}}\}| \sA):= \frac{\tvepsilon_{1}\cdot\tvepsilon_{3}}{k_{2\, 0}}\,\sA\cdot\tilde{\epsilon}_{2}(k_{1\,0}e_2-k_{2\,0}e_1)+ \mathrm{cyclic}
\end{equation}
Note that both $F$ and $\mathsf{C}$ are real functions, in the sense that they take real values provided the kinematic data is real-valued.

To double copy the integrand \eqref{dc1*}, one performs a sequence of simple steps:
\begin{enumerate}

 \item Flip the charge (i.e., the sign of the colour factor of $\sA$) to define 
 $\widetilde{\mathcal{A}}_3=F-\mathsf{C}$ and regard this as the conjugate of $\cA_{3}$:
 \be\label{dcs1}
|\cA_3|^2:= \cA_{3}\,\widetilde\cA_3=
  F^{2}(\{k_{{r}\,0},k_{{r}\,a},\tilde{\epsilon}_{{r}}\}) - \mathsf{C}^{2}(\{k_{{r}\,0},k_{{r}\,a},\tilde{\epsilon}_{{r}}\}|\sA)
 \ee
 
 \item Replace every spatial $(d-2)$-momentum 
by a curved version using the vielbein of the gravitational plane wave background (e.g., $k_{1\,a}\rightarrow k_{1\,i} E^{i}_{a}$). Replace the gauge background polarisations $\tilde{\epsilon}_a$ with gravitational background polarisations $\epsilon_a$. This yields\footnote{The latter operation is just a relabelling by removing all tildes. In particular, this replacement implies $\tvepsilon_{{r}}\cdot\tvepsilon_{{s}} \rightarrow \vepsilon_{{r}}\cdot\vepsilon_{{s}} $.}

 \be\label{dcs2}
 F^{2}(\{k_{{r}\,0},k_{{r}\,i} E^{i}_{a},\epsilon_{{r}}\}) - \mathsf{C}^{2}(\{k_{{r}\,0},k_{{r}\,i} E^{i}_{a},\epsilon_{{r}}\}|\sA)\,.
 \ee
 
 \item Replace the remaining (quadratic) dependence on the background gauge field with dependence on the background gravitational field using the rule:
 \be\label{dcs3}
 e_{r}e_{s}\,\sA^{a}\,\sA^{b} \rightarrow \left\{
 \begin{array}{c c}
   \im\, k_{{r}\,0}\,\sigma^{ab} & \mbox{if  } {r}={s} \\
   \im\, (k_{{r}\,0}+k_{{s}\,0})\,\sigma^{ab} & \mbox{otherwise}
 
 \end{array}\right.\,,
 \ee
 where $e_{{r}}$ is the charge under the background gauge field associated with external state ${r}=1,2,3$.
\end{enumerate}

The final step is motivated by dimensional considerations and suggested by the fact that $\sA_a$ encodes the gauge theory memory effect; if it is set to vanish in the in-region it will generically be a non-zero constant in the out-region remembering an integral of the field. Thus the quadratic combination $\sA_{a}\,\sA_{b}$ is where the memory effect can be seen in the amplitude.  In the gravitational case, the deformation tensor $\sigma_{ab}$ can be chosen to vanish in the past, but is then non-trivial in the future, although now  generically falling off asymptotically as  $u^{-1}$, by \eqref{newvb}. 
Therefore, the replacement \eqref{dcs3} identifies the fields responsible for memories, albeit with different functional dependence on $u$.  An additional power of momenta is needed on the gravitational side to ensure that the two combinations have the same mass dimension.    

Steps 1-3 result in an expression of the form
\be\label{dc2}
F^{2}(\{k_{{r}\,0},k_{{r}\,i} E^{i}_{a},\epsilon_{{r}}\}) - \mathsf{C}^{2}(\{k_{{r}\,0},k_{{r}\,i} E^{i}_{a},\epsilon_{{r}}\}|\sigma)\,.
\ee
Working on the support of momentum conservation in the $v$-direction -- which holds regardless of the asymptotic configuration of the three external states -- a bit of algebra reveals that
\be\label{dc3}
\mathsf{C}^{2}(\{k_{{r}\,0},k_{{r}\,i} E^{i}_{a},\epsilon_{{r}}\}|\sigma)=\im\, k_{1\,0}k_{2\,0}k_{3\,0}\,\sigma^{ab}\,\mathcal{C}_{a}\,\mathcal{C}_{b}\,,
\ee
and therefore that the expression \eqref{dc2} is in fact \emph{equal} to the 3-point integrand for gravitons on the gravitational plane wave background. 

\smallskip

There is also a canonical way to map the 3-point integrand for gluons on a gauge theory background to the 3-point integrand for gluons on a gravity background. This entails a `classical' double copy of the background (in the sense of~\cite{Monteiro:2014cda}) while leaving the functional form of the integrand unchanged. To see how this works,  use the integrand expression:
\be\label{sc1}
\cA_{3}(a_{1},a_{2},a_{3})=\tvepsilon_{1}\cdot\tvepsilon_{3}\,\sK_{1}\cdot\tvepsilon_{2}+\tvepsilon_{1}\cdot\tvepsilon_{2}\,\sK_{2}\cdot\tvepsilon_{3} +\tvepsilon_{2}\cdot\tvepsilon_{3}\,\sK_{3}\cdot\tvepsilon_{1}\,,
\ee
where $\sK_{{r}\,a}$ and $\tvepsilon_{{r}\,a}$ are given by \eqref{csmom1}, \eqref{gfpol} for ${r}=1,2,3$. Now perform the following replacements everywhere in \eqref{sc1}:
\be\label{sc2}
k_{{r}\,a}\rightarrow k_{{r}\,i}\,E^{i}_{a}\,, \qquad \tilde{\epsilon}_{{r}\,a}\rightarrow \epsilon_{{r}\,a} \,, \qquad e_{r}\,\sA_{a}\rightarrow k_{{r}\,0}\,\sigma_{ab}\,x^{b}\,.
\ee
The last of these replacements is motivated by the observation that the non-trivial component of the plane wave gauge field, namely $x^{a}\,\dot{\sA}_{a}$ is a linear function of $x$ while the non-trivial component of the plane wave metric, namely $-\ddot{E}^{i}_{a}E_{b\,i}\,x^{a} x^{b}$, is quadratic. 

After making the replacements \eqref{sc2}, the polarization vectors in the gauge field background are mapped directly onto the polarization vectors in the gravitational background: $\tvepsilon_{{r}\,\mu}\rightarrow\vepsilon_{{r}\,\mu}$. Although $\sK_{{r}\,\mu}$ is not quite mapped onto $K_{{r}\,\mu}$, it is easy to see that
\begin{equation*}
 \sK_{{r}}\cdot\tvepsilon_{{s}}\rightarrow K_{{r}}\cdot\vepsilon_{{s}}\,.
\end{equation*}
Calling this substitution map $\psi$, it follows immediately that
\be\label{sc3}
\psi\left(\cA_{3}(a_{1},a_{2},a_{3})\right)=\cM_{3}(A_1, A_2, A_3)\,,
\ee
where the two integrands have the same kinematic data but are defined on different backgrounds.


\section{Discussion}
\label{Discuss}

In this paper we have made a preliminary investigation of how the notion of double copy generalizes to curved scattering backgrounds starting with the three point amplitude on sandwich plane waves.  We find new features, but see that the double copy nevertheless does extend to this curved setting: 3-point graviton amplitudes on a plane wave space-time can be obtained by taking the double copy of 3-point gluon amplitudes on a gauge theory plane wave background.

This statement can be expressed succinctly by encoding steps 2 and 3 of the double copy procedure in  a `replacement map' $\rho$, that acts on the spaces of $(d-2)$-kinematics and background gauge fields. The double copy for 3-point integrands on plane wave backgrounds is then simply:
\be\label{dc4}
\boxed{\cM_{3}(h_1, h_2, h_3)= \rho\left(|\cA_{3}(a_1, a_2, a_3)|^2\right)\,.}
\ee
This is consistent with the usual double copy on flat backgrounds expressed in  the KLT relations. In a flat background, $\rho$ acts trivially and this is the usual squaring relation.

We have only investigated the simplest scattering amplitudes (i.e., 3-point amplitudes), which are generated by contact interactions in the space-time action. Higher-point amplitudes will involve propagator contributions; although explicit forms for propagators on plane wave backgrounds are known (e.g., \cite{Wolkow:1935zz,Ilderton:2012qe,Gibbons:1975jb,Harte:2012uw}), these are significantly more complicated that those arising from flat space. Nevertheless, the prescription given in section~\ref{TDC} seems universal: it dictates how to double copy the data for any $n$-point scattering amplitude.
Steps 1-3 do not depend on the number of external particles being three. So one can  optimistically conjecture a heuristic form of the double copy for $n$-point integrands on plane wave backgrounds:
\be\label{dc5}
\cM_{n}(h_{1},\ldots,h_{n}) = \rho\left(\sum_{\alpha,\beta\in S_{n}/\Z_{n}} \cA_{n}(\alpha)\,\mathcal{S}^{\sA}[\alpha|\beta]\,\widetilde{\cA}_{n}(\beta)\right)\,,
\ee
where the sum is over distinct colour-orderings for the $n$-point integrands on the gauge theory background, $\rho$ is the replacement map defined by steps 2 and 3 of the double copy, $\widetilde{\cA}_{n}$ is the integrand with opposite charges for the background and $\mathcal{S}^{\sA}[\alpha|\beta]$ is a plane wave analogue of the KLT matrix (perhaps obtained from the same replacement algorithm for the momenta).  However, now the $\cA$ and $\widetilde \cA$ must incorporate the non-trivial propagators on those backgrounds, and it is likely that these must also be subject to some replacement to work correctly on a gravitational background.

\smallskip

Our procedure is not a straightforward local identification of integrands.  It requires the replacement of certain structural functions appropriate for propagation on a gauge theory background by those for a gravitational background.  Indeed, colour/kinematics duality is usually expressed locally in momentum space, and so should not be expected to be local in space-time.   Here we see evidence that a non-local procedure based on  Hamilton-Jacobi functions for propagation of momentum eigenstates from null infinity will do the trick.  Thus, the most optimistic message from this for the general curved colour-kinematic duality is that although a space-time procedure cannot be local, it can work  by referring to null infinity, using Hamilton-Jacobi generating functions to create the identifications.

It would also be desirable to extend the double copy to other curved backgrounds. Although plane waves are a very special example of such backgrounds, there is some sense in which they are universal limits of \emph{all} space-times~\cite{Penrose:1976}. It would be interesting to see in what sense the results found here inform those for more general space-times.

Finally, we note that our original motivation for considering scattering on plane wave backgrounds was to provide a space-time result to compare with an alternative calculation of these amplitudes using ambitwistor string theory~\cite{Mason:2013sva} adapted to a curved background~\cite{Adamo:2014wea}. As we will show in~\cite{Adamo:2017}, ambitwistor strings provide an alternative `stringy' approach to calculating amplitudes on curved backgrounds which gives pure field theory amplitudes without $\alpha'$ corrections, in a way that cleanly manifests the double copy found here.  The use of Hamilton-Jacobi functions to bring in momenta and polarization vectors from null infinity should then tie in with the work in \cite{Adamo:2014yya,Geyer:2014lca,Adamo:2015fwa} where ambitwistor strings are formulated at null infinity.

\acknowledgments

We would like to thank Pedro Vieira, Kai R\"ohrig, David Skinner and Piotr Tourkine for useful conversations. TA, EC and LM thank the Kavli Institute for Theoretical Physics for hospitality while this work was completed; this research was supported in part by the National Science Foundation under Grant No. NSF PHY-1125915. TA is supported by an Imperial College Junior Research Fellowship; EC and LM are supported by EPSRC grant EP/M018911/1; SN is supported by EPSRC grant EP/M50659X/1 and a Studienstiftung des deutschen Volkes scholarship.


\appendix
\section{The Impulsive Plane Wave}
\label{Impulse}

For both gauge theory and gravitational sandwich plane waves, the computation of 3-point amplitudes (rather than integrands) boils down to performing integrations that depend on the particulars of the background geometry. In this appendix, we consider the simplest concrete example of a sandwich plane wave: the \emph{impulsive plane wave}~\cite{Aichelburg:1970dh,Penrose:1972xrn,Dray:1984ha,Klimcik:1988az,Ferrari:1988cc}. Impulsive plane waves correspond to gluing two flat regions together along an infinitesimal burst of radiation; in other words, the radiation region of the sandwich plane wave has delta function support. In the case of the impulsive gauge theory background, the scalar and gluon 3-point amplitudes can be computed in closed form. For the impulsive gravitational background, the 3-point amplitudes can be written in terms of integrals which are suitable to numerical approximation.

\subsection{Gauge theory background}

For an impulsive gauge theory plane wave, we have
\be\label{gimp1}
\dot{\sA}_{a}(u)=\delta(u)\,\sa_{a}\,,
\ee
for $\sa_{a}$ a set of $d-2$ constants which characterize the impulsive wave. Using the asymptotic conditions \eqref{gfbc}, it follows that
\be\label{gimp2}
\sA^{-}_{a}(u)=\Theta(u)\,\sa_{a}\,, \qquad \sA^{+}_{a}(u)=-\Theta(-u)\,\sa_{a}\,,
\ee
where $\Theta(u)$ is the Heaviside step function. Proceeding from \eqref{cs3p1} it is a straightforward calculation to obtain the 3-point amplitudes of charged scalars on this background. The results for the two independent configurations -- all incoming or two incoming and one outgoing -- are given by:
\begin{multline}\label{gimps1}
 M_{3}(\Phi^{-}_{1},\Phi^{-}_{2},\Phi^{-}_{3})=\frac{\lambda}{6}\,\delta^{d-1}\!\left(\sum_{{r}=1}^3 k_{r}\right)\left[\left(\sum_{{s}=1}^{3}\frac{\mathbf{k}_{{s}}^{2}}{2\,k_{{s}\,0}}\right)^{-1} \right. \\
 \left.-\left(\sum_{{s}=1}^{3}\frac{\mathbf{k}_{{s}}^{2}+2 e_{s} k_{{s}}^{a}\sa_{a}+e_{s}^{2}\sa^{2}}{2\,k_{{s}\,0}}\right)^{-1}\right]\,,
\end{multline}
and
\begin{multline}\label{gimps2}
 M_{3}(\Phi^{-}_{1},\Phi^{-}_{2},\Phi^{+}_{3})=\frac{\lambda}{6}\,\delta^{d-1}\!\left(\sum_{{r}=1}^3 k_{r}\right)\,\left[\left(\frac{\mathbf{k}_{3}^{2}-2e_{3} k_{3}^{a}\sa_{a}+e_{3}^{2}\sa^{2}}{2\,k_{3\,0}}+\sum_{{s}=1,2}\frac{\mathbf{k}_{{s}}^{2}}{2\,k_{{s}\,0}}\right)^{-1} \right. \\
 \left.-\left(\frac{\mathbf{k}_{3}^2}{2\,k_{0\,3}}+\sum_{{s}=1,2}\frac{\mathbf{k}_{{s}}^{2}+2e_{s} k_{{s}}^{a}\sa_{a}+e_{s}^{2}\sa^{2}}{2\,k_{{s}\,0}}\right)^{-1}\right]\,,
\end{multline}
where $\mathbf{k}_{{s}}^{2}:=k_{{s}\,a}k^{a}_{{s}}$ for any ${s}=1,2,3$.

The 3-point amplitudes for gluons on the impulsive gauge theory background follow similarly from \eqref{cgf3p4}. A calculation leads to:
\begin{multline}\label{gimpg1}
 M_{3}(a^{-}_{1},a^{-}_{2},a^{-}_{3})=2\,g\,\delta^{d-1}\!\left(\sum_{{r}=1}^3 k_{r}\right)\left[\left(\sum_{{s}=1}^{3}\frac{\mathbf{k}_{{s}}^{2}}{2\,k_{{s}\,0}}\right)^{-1}\,F(\{k_{t}, \tepsilon_{t}\}) \right. \\
 -\left(\sum_{{s}=1}^{3}\frac{\mathbf{k}_{{s}}^{2}+2e_{s} k_{{s}}^{a}\sa_{a}+e_{s}^{2}\sa^{2}}{2\,k_{{s}\,0}}\right)^{-1} \left(F(\{k_{t},\tepsilon_{t}\}) -\sa^{a} \left(\frac{\tepsilon_{1}\cdot \tepsilon_{3}}{k_{2\,0}}\,\tepsilon_{2\,a}(e_{2}k_{1\,0}-e_{1}k_{2\,0}) \right.\right. \\
 +\left.\left.\left.\frac{\tepsilon_{1}\cdot \tepsilon_{2}}{k_{3\,0}}\,\tepsilon_{3\,a}(e_{3}k_{2\,0}-e_{2}k_{3\,0}) +\frac{\tepsilon_{2}\cdot \tepsilon_{3}}{k_{1\,0}}\,\tepsilon_{1\,a} (e_{1}k_{3\,0}-e_{3}k_{1\,0})\right)\right)\right]\,,
\end{multline}
and
\begin{multline}\label{gimpg2}
 M_{3}(a^{-}_{1},a^{-}_{2},a^{+}_{3})=2\,g\,\delta^{d-1}\!\left(\sum_{{r}=1}^3 k_{r}\right)\,\left[\left(\frac{\mathbf{k}_{3}^{2}-2e_{3} k_{3}^{a}\sa_{a}+e_{3}^{2}\sa^{2}}{2\,k_{3\,0}}+\sum_{{s}=1,2}\frac{\mathbf{k}_{{s}}^{2}}{2\,k_{{s}\,0}}\right)^{-1} \right. \\
 \times\left(F(\{k_{t},\tepsilon_{t}\})+e_{3}\,\sa^{a}\left(\frac{k_{2\,0}}{k_{3\,0}}\,\tepsilon_{1}\cdot \tepsilon_{2}\,\tepsilon_{3\,a}-\tepsilon_{2}\cdot \tepsilon_{3}\,\tepsilon_{1\,a}\right)\right) \\
 -\left(\frac{\mathbf{k}_{3}^2}{2\,k_{0\,3}}+\sum_{{s}=1,2}\frac{\mathbf{k}_{{s}}^{2}+2e_{s} k_{{s}}^{a}\sa_{a}+e_{s}^{2}\sa^{2}}{2\,k_{{s}\,0}}\right)^{-1}\,\left(F(\{k_{t},\tepsilon_{t}\})-\sa^{a} \left(\frac{\tepsilon_{1}\cdot \tepsilon_{3}}{k_{2\,0}}\,\tepsilon_{2\,a}(e_{2}k_{1\,0}-e_{1}k_{2\,0})\right.\right. \\
 \left.\left.\left.-e_{2}\,\tepsilon_{1}\cdot \tepsilon_{2}\,\tepsilon_{3\,a}+e_{1}\,\frac{k_{3\,0}}{k_{1\,0}}\,\tepsilon_{2}\cdot \tepsilon_{3}\,\tepsilon_{1\,a}\right)\right)\right]\,,
\end{multline}
where the function $F$ of the kinematic data is defined by \eqref{flatamp}.

In each of these expressions a Hartle-Hawking contour deformation is used to dampen rapidly oscillating contributions to the $u$-integrations near $u=\pm\infty$. This is the same as the prescription on Minkowski space, and corresponds to selecting the physical vacuum.


\subsection{Gravitational background}

For an impulsive gravitational wave, the non-trivial metric component $H(u,\mathbf{x})$ in Brink-- mann coordinates has delta function support:
\be\label{imp1}
H(u,\mathbf{x})=\delta(u)\,H_{ab}\,x^{a}\,x^{b}\,,
\ee
with $H_{ab}$ a trace-free and constant $(d-2)\times(d-2)$ matrix. Assuming that $H_{ab}$ is corank zero with distinct eigenvalues, it can be diagonalized using rotations in the $x^{a}$-plane. So without loss of generality, we take
\be\label{imp2}
H_{ab}=\lambda_{(a)}\,\delta_{ab}\,, \qquad \sum_{a=1}^{d-2}\lambda_{(a)}=0\,.
\ee
The vielbein $E^{a}_{i}$ must solve the equation
\be\label{imp3}
\ddot{E}_{a\,i}=\lambda_{(a)}\,\delta_{ab}\,\delta(u)\,E^{b}_{i}\,,
\ee
subject to incoming or outgoing boundary conditions \eqref{vbbc}. In each case, one finds
\be\label{imp4}
E^{-}_{a\,i}=\delta_{ai}\left(1+u\,\lambda_{(a)}\,\Theta(u)\right)\,, \qquad E^{+}_{a\,i}=\delta_{ai}\left(1-u\,\lambda_{(a)}\,\Theta(-u)\right)\,,
\ee
so the transverse metric $\gamma_{ij}(u)$ is given in incoming or outgoing coordinates by:
\be\label{impmet}
\gamma_{ij}^{-}(u)=\delta_{ij}\left(1+u\,\lambda_{(i)}\,\Theta(u)\right)^{2}\,, \qquad \gamma_{ij}^{+}(u)=\delta_{ij}\left(1-u\,\lambda_{(i)}\,\Theta(-u)\right)^{2}\,,
\ee
where $\lambda_{(i)}$ is identified with $\lambda_{(a)}$ using $\delta_{a}^{i}$. This demonstrates that the impulsive gravitational wave is two copies of Minkowski space glued together along a single pulse of gravitational radiation. While the metrics \eqref{impmet} are continuous across the pulse at $u=0$, they have discontinuous first derivatives.

To compute 3-point amplitudes, it is also important to have the inverse vielbeins:
\be\label{imp5}
E^{i\,-}_{a}=\delta^{i}_{a}\left(1+u\,\lambda_{(a)}\,\Theta(u)\right)^{-1}\,, \qquad E^{i\,+}_{a}=\delta^{i}_{a}\left(1-u\,\lambda_{(a)}\,\Theta(-u)\right)^{-1}\,,
\ee
leading to expressions for $F^{ij}_{\pm}(u)$:
\begin{subequations}\label{imp6}
 \be
 F^{ij}_{-}(u)=\frac{u\,\delta^{ij}}{1+u\,\lambda_{(i)}\,\Theta(u)},
 \ee
 \be
 F^{ij}_{+}(u)=\frac{u\,\delta^{ij}}{1-u\,\lambda_{(i)}\,\Theta(-u)}\,.
 \ee
\end{subequations}
So in both cases $F^{ij}(u)$ gets an infinite series of $O(u^2)$ corrections upon crossing the pulse at $u=0$.

Even at the level of scalar amplitudes, the situation on the gravitational background is more complicated than on the gauge theory background. Unlike \eqref{gimps1}--\eqref{gimps2}, on the impulsive gravitational wave (relatively) compact expressions for the $u$-integrations are not available. Instead, we find explicit expressions which could be evaluated (numerically or possibly analytically) when the momenta and eigenvalues $\{\lambda_{(a)}\}$ are specified. For instance, one finds:
\begin{multline}\label{imps1}
 M_{3}(\Phi^{-}_{1},\Phi_{2}^{-},\Phi^{-}_{3})=\frac{\lambda\,\im}{6}\,\delta^{d-1}\!\left(\sum_{{r}=1}^{3}k_{r}\right)\,\left[\left(\sum_{{s}=1}^{3}\frac{\mathbf{k}_{{s}}^{2}}{2\,k_{{s}\,0}}\right)^{-1} \right. \\
 \left. +\im\int\limits_{0}^{\infty+i\epsilon}\d u\,\prod_{a=1}^{d-2}(1+u\,\lambda_{(a)})^{-\frac{1}{2}}\,\exp\left(\im\,u \sum_{{s}=1}^{3}\sum_{i=1}^{d-2}\frac{k_{s\,i}^{2}}{2k_{s\,0}\,(1+u\lambda_{(i)})}\right)\right]\,,
\end{multline}
for the all-incoming configuration.

The expression for the two-incoming, one-outgoing configuration is similarly given in terms of $u$-integrals over the in- and out-regions:
\begin{multline}\label{imps2}
 M_{3}(\Phi^{-}_{1},\Phi_{2}^{-},\Phi^{+}_{3})=-\frac{\lambda}{6}\,\sqrt{\frac{(2\,\pi\im)^{d-2}}{k_{3\,0}^{d-2}}}\;\delta\!\left(\sum_{{r}=1}^{3}k_{{r}\,0}\right) \\
 \times\left[\int\limits_{-\infty-\im\epsilon}^{0}\frac{\d u}{\prod_{a=1}^{d-2}\sqrt{\lambda_{(a)}}}\,\exp\left(-\frac{\im}{2\,k_{3\,0}}J_{a}J_{b} (A^{-1})^{ab}+\im\sum_{{s}=1}^{3}\frac{k_{{s}\,i}k_{{s}\,j}}{2\,k_{{s}\,0}} F^{ij}_{{s}}\right)\right. \\
 \left.+\int\limits^{\infty+\im\epsilon}_{0} \d u\,\prod_{a=1}^{d-2}(\lambda_{(a)}+u\,\lambda^{2}_{(a)})^{-\frac{1}{2}}\,\exp\left(-\frac{\im}{2\,k_{3\,0}}J_{a}J_{b} (A^{-1})^{ab}+\im\sum_{{s}=1}^{3}\frac{k_{{s}\,i}k_{{s}\,j}}{2\,k_{{s}\,0}} F^{ij}_{{s}}\right)\right]\,.
\end{multline}
Here, the $F^{ij}_{{s}}(u)$ are given by \eqref{imp6}, while
\be\label{imp7}
A_{ab}(u)=\frac{-\lambda_{(a)}\,\delta_{ab}}{1+|u|\,\lambda_{(a)}}\,,
\ee
and
\be\label{imp8}
J_{a}(u)=\frac{k_{1\,a}+k_{2\,a}+k_{3\,a}+u\,\lambda_{(a)}(k_{3\,a}\,\Theta(u)-(k_{1\,a}+k_{2\,a})\,\Theta(-u))}{1+|u|\,\lambda_{(a)}}\,.
\ee

\section{Classical S-matrix \& Tree-level integrands}
\label{S-matrix-integrands}

This appendix reviews the notion of classical S-matrix which is used throughout the paper, as well as providing a precise definition for the \emph{tree-level integrand}. On a sandwich plane wave background (for either gauge theory or gravity), the tree-level S-matrix for a theory encodes the evolution of asymptotic free states from the in-region of the space-time (i.e., $u<u_1$) through the non-trivial, or radiation region ($u_{1}\leq u\leq u_{2}$), to the out-region ($u>u_2$) as governed by the classical theory. 


Rather than work out the curved space Feynman rules, we use a definition of the classical S-matrix in which tree-level amplitudes are given by extracting certain multi-linear pieces of the classical action evaluated on a perturbative solution to the non-linear equations~\cite{Arefeva:1974jv,Jevicki:1987ax,Rosly:1996vr}.  In general this has the interpretation of the field-theoretic Hamilton-Jacobi generating function for the evolution and gives the tree-level contribution to the S-matrix. For the 3-point calculations in the body of the paper, there is no need to iterate the perturbative solution, but here we present the general framework.   

Let $S[\Phi]$ be the classical action, a functional of some fields $\Phi$ which is defined on the sandwich plane wave background (gravitational or gauge theoretic -- at this stage it makes no difference). We assume that this action takes the generic form:
\be\label{smat1}
S[\Phi]=\int \d^{d}X \left(\cL_{\mathrm{kin}}+\cL_{\mathrm{int}}\right)\,,
\ee
where $\cL_{\mathrm{kin}}$ is the kinetic portion of the action, which is quadratic in $\Phi$ and governs the free theory, and $\cL_{\mathrm{int}}$ contains all higher-point interactions.

Define the following object:
\be\label{smat2}
\Phi^{[n]}(X):=\sum_{i=1}^{n}\epsilon_{i}\,\varphi_{i}(X)+\int \d^{d}Y\,\Delta(X,Y)\,\left.\frac{\delta\cL_{\mathrm{int}}}{\delta\Phi}\right|_{\Phi=\sum_{j=1}^{n}\epsilon_{j}\varphi_{j}(Y)}.
\ee
This is essentially an integral form of the full non-linear equations from the action $S$ with data given by the first term on the right hand side. Here, the $\{\epsilon_{i}\}$ are $n$ parameters that will eventually  be thought of as infinitesimal; $\{\varphi_{i}\}$ are $n$ solutions to the free equations of motion of $\cL_{\mathrm{kin}}$ with specified asymptotic behaviour; and $\Delta(X,Y)$ is a Green's function defined by $\cL_{\mathrm{kin}}$. There are precise formulae for various useful definitions of this $\Delta(X,Y)$ (e.g., advanced, retarded, Feynman) in scalar, gauge, and gravitational theories on plane wave backgrounds~\cite{Gibbons:1975jb,Harte:2012uw}, though we will not make explicit use of them here. Specifying the asymptotic behaviour of the free solution $\varphi_{i}$ boils down to saying whether it looks like an `in' or `out' state.

Both the in- and out-regions are flat, so asymptotically free states $\varphi_{i}$ should look like free states in Minkowski space in at least one of these regions. In a momentum space representation, such free states in Minkowski space are modelled on massless plane wave momentum eigenstates, $\e^{\im\,k\cdot X}$ for $k^2=0$. Unlike Minkowski space, in the sandwich plane wave a state which looks like $\e^{\im\,k\cdot X}$ in the in-region will \emph{not} look like $\e^{\im\,k\cdot X}$ in the out-region. This is a consequence of the `memory' relations \eqref{newvb}, \eqref{gmem}. Hence, the specification of asymptotic behaviour for $\varphi_{i}$ boils down to stating whether it is an \emph{incoming} or \emph{outgoing} state, denoted respectively as $\varphi_{i}^{-}$ or $\varphi_{i}^{+}$. An incoming state is one which looks like a free solution in Minkowski space the in-region; an out state looks like a free solution in Minkowski space in the out-region. More precisely,
\be\label{smat3} 
\varphi^{-}_{i}|_{\mathrm{in}}\sim\e^{\im\,k\cdot X}\sim\varphi^{+}_{i}|_{\mathrm{out}}\,,
\ee
for both the gravitational and gauge theory backgrounds.

The $n$-point tree-level scattering amplitude for the states $\{\varphi_{i}\}$ -- with their given asymptotic configuration of in and out states -- is then a multi-linear piece of the classical action:
\be\label{smat4}
M^{(0)}_{n}(\varphi_{1},\ldots,\varphi_{n})=\left.\frac{\partial^{n}S[\Phi^{[n]}]}{\partial\epsilon_{1}\cdots\partial\epsilon_{n}}\right|_{\epsilon_{1}=\cdots=\epsilon_{n}=0}\,.
\ee
For flat backgrounds, this agrees with the usual definition of the S-matrix and would also correspond with a Feynman diagram definition for sandwich plane waves.

\medskip

For the purposes of investigating the double copy, a notion of \emph{tree-level integrand} closely related to the tree-level amplitude is useful. Indeed, it is actually this tree-level integrand that appears in the KLT relations of the standard double copy. From the definition \eqref{smat4} it is straightforward to see that the tree-level scattering amplitude will always take the form:
\be\label{tint1}
M^{(0)}_{n} = \int \d^{d}X\,\cM_{n}(X)\,\prod_{i=1}^{n} f_{i}(X)\,,
\ee
where each of the $f_{i}(X)$ is a solution to the free scalar wave equation on the plane wave background. The object $\cM_{n}$ is defined to be the tree-level integrand; generically, it will be formed of polarizations, momenta and propagators and depends on the background geometry. It captures everything that is encoded by the kinematic numerators and denominators which would result from a conventional Feynman diagram approach. In more heuristic terms, the tree-level integrand is what remains after removing the final integral that forms the action functional in \eqref{smat4}, along with `universal' spin-independent functions.  

In Minkowski space, it is easy to see that
\begin{equation*}
\prod_{i=1}^{n} f_{i}(X)=\e^{\im(k_{1}+\cdots+k_{n})\cdot X}\,,
\end{equation*}
so the effect of isolating $\cM_{n}$ is to strip off an overall momentum conserving delta function. On non-trivial backgrounds such as the sandwich plane wave, the result of the final $\d^{d}X$ integrals is more complicated, but the principle is the same: $\cM_{n}$ contains all of the information which one could expect to be `squared' in taking the double copy. Another interesting property of the integrand is that it is functionally independent of the asymptotic conditions of the states being scattered. This enables the investigation of double copy by considering the computationally simplest configuration of incoming and outgoing states.

Clearly, there is a sense in which the tree-level integrand is not a gauge-invariant object, just as one can add boundary terms to an action. This lack of gauge invariance is analogous to the statement that individual Feynman diagrams -- or individual terms contributing to \eqref{smat4} -- are not gauge invariant. However, once a gauge for performing perturbative calculations has been fixed (i.e., specific linearized solutions $\{\varphi_i\}$ and a Green's function $\Delta(X,Y)$ have been consistently chosen), the object $\cM_{n}$ is well-defined. In our calculations, we always work in a Lorenz or de Donder gauge, so the resulting expressions for the integrand should be viewed as expressions in these particular gauges. Their \emph{integrals}, however, do not depend on the gauge choice.

Throughout the paper, the tree-level integrand for theories on the gravitational plane wave background is denoted by $\cM_{n}$, and the tree-level integrand for theories on the gauge theory plane wave background by $\cA_{n}$.


\bibliography{planwaves}

\providecommand{\href}[2]{#2}\begingroup\raggedright\begin{thebibliography}{10}

\bibitem{Kawai:1985xq}
H.~Kawai, D.~C. Lewellen and S.~H.~H. Tye, \emph{{A Relation Between Tree
  Amplitudes of Closed and Open Strings}},
  \href{http://dx.doi.org/10.1016/0550-3213(86)90362-7}{\emph{Nucl. Phys.}
  {\bfseries B269} (1986) 1--23}.

\bibitem{Bern:2008qj}
Z.~Bern, J.~J.~M. Carrasco and H.~Johansson, \emph{{New Relations for
  Gauge-Theory Amplitudes}},
  \href{http://dx.doi.org/10.1103/PhysRevD.78.085011}{\emph{Phys. Rev.}
  {\bfseries D78} (2008) 085011},
  [\href{https://arxiv.org/abs/0805.3993}{{\ttfamily 0805.3993}}].

\bibitem{Bern:2010ue}
Z.~Bern, J.~J.~M. Carrasco and H.~Johansson, \emph{{Perturbative Quantum
  Gravity as a Double Copy of Gauge Theory}},
  \href{http://dx.doi.org/10.1103/PhysRevLett.105.061602}{\emph{Phys. Rev.
  Lett.} {\bfseries 105} (2010) 061602},
  [\href{https://arxiv.org/abs/1004.0476}{{\ttfamily 1004.0476}}].

\bibitem{Bern:2010yg}
Z.~Bern, T.~Dennen, Y.-t. Huang and M.~Kiermaier, \emph{{Gravity as the Square
  of Gauge Theory}},
  \href{http://dx.doi.org/10.1103/PhysRevD.82.065003}{\emph{Phys. Rev.}
  {\bfseries D82} (2010) 065003},
  [\href{https://arxiv.org/abs/1004.0693}{{\ttfamily 1004.0693}}].

\bibitem{Bern:2009kd}
Z.~Bern, J.~J. Carrasco, L.~J. Dixon, H.~Johansson and R.~Roiban, \emph{{The
  Ultraviolet Behavior of N=8 Supergravity at Four Loops}},
  \href{http://dx.doi.org/10.1103/PhysRevLett.103.081301}{\emph{Phys. Rev.
  Lett.} {\bfseries 103} (2009) 081301},
  [\href{https://arxiv.org/abs/0905.2326}{{\ttfamily 0905.2326}}].

\bibitem{Bern:2012cd}
Z.~Bern, S.~Davies, T.~Dennen and Y.-t. Huang, \emph{{Absence of Three-Loop
  Four-Point Divergences in N=4 Supergravity}},
  \href{http://dx.doi.org/10.1103/PhysRevLett.108.201301}{\emph{Phys. Rev.
  Lett.} {\bfseries 108} (2012) 201301},
  [\href{https://arxiv.org/abs/1202.3423}{{\ttfamily 1202.3423}}].

\bibitem{Bern:2012gh}
Z.~Bern, S.~Davies, T.~Dennen and Y.-t. Huang, \emph{{Ultraviolet Cancellations
  in Half-Maximal Supergravity as a Consequence of the Double-Copy Structure}},
  \href{http://dx.doi.org/10.1103/PhysRevD.86.105014}{\emph{Phys. Rev.}
  {\bfseries D86} (2012) 105014},
  [\href{https://arxiv.org/abs/1209.2472}{{\ttfamily 1209.2472}}].

\bibitem{Bern:2013uka}
Z.~Bern, S.~Davies, T.~Dennen, A.~V. Smirnov and V.~A. Smirnov,
  \emph{{Ultraviolet Properties of N=4 Supergravity at Four Loops}},
  \href{http://dx.doi.org/10.1103/PhysRevLett.111.231302}{\emph{Phys. Rev.
  Lett.} {\bfseries 111} (2013) 231302},
  [\href{https://arxiv.org/abs/1309.2498}{{\ttfamily 1309.2498}}].

\bibitem{Bern:2014sna}
Z.~Bern, S.~Davies and T.~Dennen, \emph{{Enhanced ultraviolet cancellations in
  $\mathcal N=5$ supergravity at four loops}},
  \href{http://dx.doi.org/10.1103/PhysRevD.90.105011}{\emph{Phys. Rev.}
  {\bfseries D90} (2014) 105011},
  [\href{https://arxiv.org/abs/1409.3089}{{\ttfamily 1409.3089}}].

\bibitem{Bern:2015xsa}
Z.~Bern, C.~Cheung, H.-H. Chi, S.~Davies, L.~Dixon and J.~Nohle,
  \emph{{Evanescent Effects Can Alter Ultraviolet Divergences in Quantum
  Gravity without Physical Consequences}},
  \href{http://dx.doi.org/10.1103/PhysRevLett.115.211301}{\emph{Phys. Rev.
  Lett.} {\bfseries 115} (2015) 211301},
  [\href{https://arxiv.org/abs/1507.06118}{{\ttfamily 1507.06118}}].

\bibitem{Bern:2017lpv}
Z.~Bern, M.~Enciso, J.~Parra-Martinez and M.~Zeng, \emph{{Manifesting enhanced
  cancellations in supergravity: integrands versus integrals}},
  \href{https://arxiv.org/abs/1703.08927}{{\ttfamily 1703.08927}}.

\bibitem{Bern:2017tuc}
Z.~Bern, A.~Edison, D.~Kosower and J.~Parra-Martinez, \emph{{Curvature-Squared
  Multiplets, Evanescent Effects and the U(1) Anomaly in N = 4 Supergravity}},
  \href{https://arxiv.org/abs/1706.01486}{{\ttfamily 1706.01486}}.

\bibitem{Bern:2006kd}
Z.~Bern, L.~J. Dixon and R.~Roiban, \emph{{Is N = 8 supergravity ultraviolet
  finite?}},
  \href{http://dx.doi.org/10.1016/j.physletb.2006.11.030}{\emph{Phys. Lett.}
  {\bfseries B644} (2007) 265--271},
  [\href{https://arxiv.org/abs/hep-th/0611086}{{\ttfamily hep-th/0611086}}].

\bibitem{Bern:2011qn}
Z.~Bern, J.~J. Carrasco, L.~J. Dixon, H.~Johansson and R.~Roiban,
  \emph{{Amplitudes and Ultraviolet Behavior of N = 8 Supergravity}},
  \href{http://dx.doi.org/10.1002/prop.201100037}{\emph{Fortsch. Phys.}
  {\bfseries 59} (2011) 561--578},
  [\href{https://arxiv.org/abs/1103.1848}{{\ttfamily 1103.1848}}].

\bibitem{BjerrumBohr:2009rd}
N.~E.~J. Bjerrum-Bohr, P.~H. Damgaard and P.~Vanhove, \emph{{Minimal Basis for
  Gauge Theory Amplitudes}},
  \href{http://dx.doi.org/10.1103/PhysRevLett.103.161602}{\emph{Phys. Rev.
  Lett.} {\bfseries 103} (2009) 161602},
  [\href{https://arxiv.org/abs/0907.1425}{{\ttfamily 0907.1425}}].

\bibitem{Stieberger:2009hq}
S.~Stieberger, \emph{{Open \& Closed vs. Pure Open String Disk Amplitudes}},
  \href{https://arxiv.org/abs/0907.2211}{{\ttfamily 0907.2211}}.

\bibitem{BjerrumBohr:2010zs}
N.~E.~J. Bjerrum-Bohr, P.~H. Damgaard, T.~Sondergaard and P.~Vanhove,
  \emph{{Monodromy and Jacobi-like Relations for Color-Ordered Amplitudes}},
  \href{http://dx.doi.org/10.1007/JHEP06(2010)003}{\emph{JHEP} {\bfseries 06}
  (2010) 003}, [\href{https://arxiv.org/abs/1003.2403}{{\ttfamily 1003.2403}}].

\bibitem{Feng:2010my}
B.~Feng, R.~Huang and Y.~Jia, \emph{{Gauge Amplitude Identities by On-shell
  Recursion Relation in S-matrix Program}},
  \href{http://dx.doi.org/10.1016/j.physletb.2010.11.011}{\emph{Phys. Lett.}
  {\bfseries B695} (2011) 350--353},
  [\href{https://arxiv.org/abs/1004.3417}{{\ttfamily 1004.3417}}].

\bibitem{Tye:2010dd}
S.~H. Henry~Tye and Y.~Zhang, \emph{{Dual Identities inside the Gluon and the
  Graviton Scattering Amplitudes}},
  \href{http://dx.doi.org/10.1007/JHEP06(2010)071,
  10.1007/JHEP04(2011)114}{\emph{JHEP} {\bfseries 06} (2010) 071},
  [\href{https://arxiv.org/abs/1003.1732}{{\ttfamily 1003.1732}}].

\bibitem{Anastasiou:2013hba}
A.~Anastasiou, L.~Borsten, M.~J. Duff, L.~J. Hughes and S.~Nagy, \emph{{A magic
  pyramid of supergravities}},
  \href{http://dx.doi.org/10.1007/JHEP04(2014)178}{\emph{JHEP} {\bfseries 04}
  (2014) 178}, [\href{https://arxiv.org/abs/1312.6523}{{\ttfamily 1312.6523}}].

\bibitem{Monteiro:2014cda}
R.~Monteiro, D.~O'Connell and C.~D. White, \emph{{Black holes and the double
  copy}}, \href{http://dx.doi.org/10.1007/JHEP12(2014)056}{\emph{JHEP}
  {\bfseries 12} (2014) 056},
  [\href{https://arxiv.org/abs/1410.0239}{{\ttfamily 1410.0239}}].

\bibitem{Luna:2015paa}
A.~Luna, R.~Monteiro, D.~O'Connell and C.~D. White, \emph{{The classical double
  copy for Taub–NUT spacetime}},
  \href{http://dx.doi.org/10.1016/j.physletb.2015.09.021}{\emph{Phys. Lett.}
  {\bfseries B750} (2015) 272--277},
  [\href{https://arxiv.org/abs/1507.01869}{{\ttfamily 1507.01869}}].

\bibitem{Ridgway:2015fdl}
A.~K. Ridgway and M.~B. Wise, \emph{{Static Spherically Symmetric Kerr-Schild
  Metrics and Implications for the Classical Double Copy}},
  \href{http://dx.doi.org/10.1103/PhysRevD.94.044023}{\emph{Phys. Rev.}
  {\bfseries D94} (2016) 044023},
  [\href{https://arxiv.org/abs/1512.02243}{{\ttfamily 1512.02243}}].

\bibitem{Borsten:2015pla}
L.~Borsten and M.~J. Duff, \emph{{Gravity as the square of Yang–Mills?}},
  \href{http://dx.doi.org/10.1088/0031-8949/90/10/108012}{\emph{Phys. Scripta}
  {\bfseries 90} (2015) 108012},
  [\href{https://arxiv.org/abs/1602.08267}{{\ttfamily 1602.08267}}].

\bibitem{Luna:2016due}
A.~Luna, R.~Monteiro, I.~Nicholson, D.~O'Connell and C.~D. White, \emph{{The
  double copy: Bremsstrahlung and accelerating black holes}},
  \href{http://dx.doi.org/10.1007/JHEP06(2016)023}{\emph{JHEP} {\bfseries 06}
  (2016) 023}, [\href{https://arxiv.org/abs/1603.05737}{{\ttfamily
  1603.05737}}].

\bibitem{Goldberger:2016iau}
W.~D. Goldberger and A.~K. Ridgway, \emph{{Radiation and the classical double
  copy for color charges}},  \href{https://arxiv.org/abs/1611.03493}{{\ttfamily
  1611.03493}}.

\bibitem{Cardoso:2016amd}
G.~Cardoso, S.~Nagy and S.~Nampuri, \emph{{Multi-centered $ \mathcal{N}=2 $ BPS
  black holes: a double copy description}},
  \href{http://dx.doi.org/10.1007/JHEP04(2017)037}{\emph{JHEP} {\bfseries 04}
  (2017) 037}, [\href{https://arxiv.org/abs/1611.04409}{{\ttfamily
  1611.04409}}].

\bibitem{Luna:2016hge}
A.~Luna, R.~Monteiro, I.~Nicholson, A.~Ochirov, D.~O'Connell, N.~Westerberg
  et~al., \emph{{Perturbative spacetimes from Yang-Mills theory}},
  \href{http://dx.doi.org/10.1007/JHEP04(2017)069}{\emph{JHEP} {\bfseries 04}
  (2017) 069}, [\href{https://arxiv.org/abs/1611.07508}{{\ttfamily
  1611.07508}}].

\bibitem{Goldberger:2017frp}
W.~D. Goldberger, S.~G. Prabhu and J.~O. Thompson, \emph{{Classical gluon and
  graviton radiation from the bi-adjoint scalar double copy}},
  \href{https://arxiv.org/abs/1705.09263}{{\ttfamily 1705.09263}}.

\bibitem{Bondi:1958aj}
H.~Bondi, F.~A.~E. Pirani and I.~Robinson, \emph{{Gravitational waves in
  general relativity. 3. Exact plane waves}},
  \href{http://dx.doi.org/10.1098/rspa.1959.0124}{\emph{Proc. Roy. Soc. Lond.}
  {\bfseries A251} (1959) 519--533}.

\bibitem{Penrose:1965rx}
R.~Penrose, \emph{{A Remarkable property of plane waves in general
  relativity}}, \href{http://dx.doi.org/10.1103/RevModPhys.37.215}{\emph{Rev.
  Mod. Phys.} {\bfseries 37} (1965) 215--220}.

\bibitem{Friedlander:1975eqa}
F.~G. Friedlander, \emph{{The Wave Equation on a Curved Space-Time}}.
\newblock Cambridge University Press, 1975.

\bibitem{Mason:1989}
L.~J. Mason, \emph{{On Ward's integral formula for the wave equation in plane
  wave space-times}}, {\emph{Twistor Newsletter} {\bfseries 28} (1989) 17--19}.

\bibitem{Braginsky:1986ia}
V.~B. Braginsky and L.~P. Grishchuk, \emph{{Kinematic Resonance and Memory
  Effect in Free Mass Gravitational Antennas}}, {\emph{Sov. Phys. JETP}
  {\bfseries 62} (1985) 427--430}.

\bibitem{Braginsky:1987}
V.~B. Braginsky and K.~S. Thorne, \emph{{Gravitational-wave bursts with memory
  and experimental prospects}}, {\emph{Nature} {\bfseries 327} (1987)
  123--125}.

\bibitem{Ludvigsen:1989kg}
M.~Ludvigsen, \emph{{Geodesic deviation at null infinity and the physical
  effects of very long wave gravitational radiation}},
  \href{http://dx.doi.org/10.1007/BF00763308}{\emph{Gen. Rel. Grav.} {\bfseries
  21} (1989) 1205--1212}.

\bibitem{Zhang:2017rno}
P.~M. Zhang, C.~Duval, G.~W. Gibbons and P.~A. Horvathy, \emph{{The Memory
  Effect for Plane Gravitational Waves}},
  \href{https://arxiv.org/abs/1704.05997}{{\ttfamily 1704.05997}}.

\bibitem{Zhang:2017geq}
P.~M. Zhang, C.~Duval, G.~W. Gibbons and P.~A. Horvathy, \emph{{Soft Gravitons
  \& the Memory Effect for Plane Gravitational Waves}},
  \href{https://arxiv.org/abs/1705.01378}{{\ttfamily 1705.01378}}.

\bibitem{Dinu:2012tj}
V.~Dinu, T.~Heinzl and A.~Ilderton, \emph{{Infra-Red Divergences in Plane Wave
  Backgrounds}},
  \href{http://dx.doi.org/10.1103/PhysRevD.86.085037}{\emph{Phys. Rev.}
  {\bfseries D86} (2012) 085037},
  [\href{https://arxiv.org/abs/1206.3957}{{\ttfamily 1206.3957}}].

\bibitem{Ilderton:2012qe}
A.~Ilderton and G.~Torgrimsson, \emph{{Scattering in plane-wave backgrounds:
  infra-red effects and pole structure}},
  \href{http://dx.doi.org/10.1103/PhysRevD.87.085040}{\emph{Phys. Rev.}
  {\bfseries D87} (2013) 085040},
  [\href{https://arxiv.org/abs/1210.6840}{{\ttfamily 1210.6840}}].

\bibitem{Baldwin:1926}
O.~R. Baldwin and G.~B. Jeffery, \emph{{The relativity theory of plane waves}},
  {\emph{Proc.Roy.Soc.Lond.} {\bfseries A111} (1926) 95}.

\bibitem{Ehlers:1962zz}
J.~Ehlers and W.~Kundt, \emph{{Exact solutions of the gravitational field
  equations}},  in \emph{{Gravitation, An Introduction to Current Research}}
  (L.~Witten, ed.), p.~49.
\newblock Wiley: New York, 1962.

\bibitem{griffiths1991colliding}
J.~Griffiths, \emph{Colliding plane waves in general relativity}.
\newblock Oxford mathematical monographs. Clarendon Press, 1991.

\bibitem{Stephani:2003tm}
H.~Stephani, D.~Kramer, M.~A.~H. MacCallum, C.~Hoenselaers and E.~Herlt,
  \emph{{Exact solutions of Einstein's field equations}}.
\newblock Cambridge University Press, 2~ed., 2004.

\bibitem{Blau:2011}
M.~Blau, \emph{{Plane waves and Penrose limits}},  tech. rep., Universit\'{e}
  de Neuch\^{a}tel, 2011.

\bibitem{Einstein:1937qu}
A.~Einstein and N.~Rosen, \emph{{On Gravitational waves}},
  \href{http://dx.doi.org/10.1016/S0016-0032(37)90583-0}{\emph{J. Franklin
  Inst.} {\bfseries 223} (1937) 43--54}.

\bibitem{Brinkmann:1925fr}
H.~W. Brinkmann, \emph{{Einstein spapces which are mapped conformally on each
  other}}, \href{http://dx.doi.org/10.1007/BF01208647}{\emph{Math. Ann.}
  {\bfseries 94} (1925) 119--145}.

\bibitem{Bondi:1989vm}
H.~Bondi and F.~A.~E. Pirani, \emph{{Gravitational Waves in General Relativity.
  13: Caustic Property of Plane Waves}},
  \href{http://dx.doi.org/10.1098/rspa.1989.0016}{\emph{Proc. Roy. Soc. Lond.}
  {\bfseries A421} (1989) 395--410}.

\bibitem{Reiss:1962}
H.~R. Reiss, \emph{{Absorption of light by light}},
  \href{http://dx.doi.org/10.1063/1.1703787}{\emph{J.Math.Phys.} {\bfseries 3}
  (1962) 59--67}.

\bibitem{Brown:1964zzb}
L.~S. Brown and T.~W.~B. Kibble, \emph{{Interaction of Intense Laser Beams with
  Electrons}}, \href{http://dx.doi.org/10.1103/PhysRev.133.A705}{\emph{Phys.
  Rev.} {\bfseries 133} (1964) A705--A719}.

\bibitem{Coleman:1977ps}
S.~R. Coleman, \emph{{Nonabelian Plane Waves}},
  \href{http://dx.doi.org/10.1016/0370-2693(77)90344-6}{\emph{Phys. Lett.}
  {\bfseries B70} (1977) 59--60}.

\bibitem{Bieri:2013hqa}
L.~Bieri and D.~Garfinkle, \emph{{An electromagnetic analogue of gravitational
  wave memory}},
  \href{http://dx.doi.org/10.1088/0264-9381/30/19/195009}{\emph{Class. Quant.
  Grav.} {\bfseries 30} (2013) 195009},
  [\href{https://arxiv.org/abs/1307.5098}{{\ttfamily 1307.5098}}].

\bibitem{Ward:1987ws}
R.~S. Ward, \emph{{Progressing waves in flat space-time and in plane wave
  space-times}},
  \href{http://dx.doi.org/10.1088/0264-9381/4/3/034}{\emph{Class. Quant. Grav.}
  {\bfseries 4} (1987) 775--778}.

\bibitem{Gibbons:1975jb}
G.~W. Gibbons, \emph{{Quantized Fields Propagating in Plane Wave Space-Times}},
  \href{http://dx.doi.org/10.1007/BF01629249}{\emph{Commun. Math. Phys.}
  {\bfseries 45} (1975) 191--202}.

\bibitem{Garriga:1990dp}
J.~Garriga and E.~Verdaguer, \emph{{Scattering of quantum particles by
  gravitational plane waves}},
  \href{http://dx.doi.org/10.1103/PhysRevD.43.391}{\emph{Phys. Rev.} {\bfseries
  D43} (1991) 391--401}.

\bibitem{Crnkovic:1986ex}
C.~Crnkovic and E.~Witten, \emph{{Covariant description of canonical formalism
  in geometrical theories}},  in \emph{{Three Hundred Years of Gravitation}}
  (S.~Hawking and W.~Israel, eds.), p.~676.
\newblock Cambridge University Press, 1986.

\bibitem{Gunther:1974}
P.~Günther and V.~Wünsch, \emph{{Maxwellsche Gleichungen und Huygenssches
  Prinzip I}}, {\emph{Mathematische Nachrichten} {\bfseries 63} (1974)
  97--121}.

\bibitem{Gunther:1988}
P.~Günther, \emph{{Huygen's Principle and Hyperbolic Differential Equations}}.
\newblock Academic Press, San Diego, 1988.

\bibitem{Harte:2013dba}
A.~I. Harte, \emph{{Tails of plane wave spacetimes: Wave-wave scattering in
  general relativity}},
  \href{http://dx.doi.org/10.1103/PhysRevD.88.084059}{\emph{Phys. Rev.}
  {\bfseries D88} (2013) 084059},
  [\href{https://arxiv.org/abs/1309.5020}{{\ttfamily 1309.5020}}].

\bibitem{Papadopoulos:2002bg}
G.~Papadopoulos, J.~G. Russo and A.~A. Tseytlin, \emph{{Solvable model of
  strings in a time dependent plane wave background}},
  \href{http://dx.doi.org/10.1088/0264-9381/20/5/313}{\emph{Class. Quant.
  Grav.} {\bfseries 20} (2003) 969--1016},
  [\href{https://arxiv.org/abs/hep-th/0211289}{{\ttfamily hep-th/0211289}}].

\bibitem{Cheung:2016say}
C.~Cheung and G.~N. Remmen, \emph{{Twofold Symmetries of the Pure Gravity
  Action}}, \href{http://dx.doi.org/10.1007/JHEP01(2017)104}{\emph{JHEP}
  {\bfseries 01} (2017) 104},
  [\href{https://arxiv.org/abs/1612.03927}{{\ttfamily 1612.03927}}].

\bibitem{Wolkow:1935zz}
D.~M. Wolkow, \emph{{Uber eine Klasse von Losungen der Diracschen Gleichung}},
  \href{http://dx.doi.org/10.1007/BF01331022}{\emph{Z. Phys.} {\bfseries 94}
  (1935) 250--260}.

\bibitem{Harte:2012uw}
A.~I. Harte and T.~D. Drivas, \emph{{Caustics and wave propagation in curved
  spacetimes}}, \href{http://dx.doi.org/10.1103/PhysRevD.85.124039}{\emph{Phys.
  Rev.} {\bfseries D85} (2012) 124039},
  [\href{https://arxiv.org/abs/1202.0540}{{\ttfamily 1202.0540}}].

\bibitem{Penrose:1976}
R.~Penrose, \emph{{Any geometry has a plane-wave limit}},  in
  \emph{{Differential Geometry and Relativity}} (M.~Cahen and M.~Flato, eds.),
  p.~271.
\newblock Reidel: Dordrecht, 1976.

\bibitem{Mason:2013sva}
L.~Mason and D.~Skinner, \emph{{Ambitwistor strings and the scattering
  equations}}, \href{http://dx.doi.org/10.1007/JHEP07(2014)048}{\emph{JHEP}
  {\bfseries 07} (2014) 048},
  [\href{https://arxiv.org/abs/1311.2564}{{\ttfamily 1311.2564}}].

\bibitem{Adamo:2014wea}
T.~Adamo, E.~Casali and D.~Skinner, \emph{{A Worldsheet Theory for
  Supergravity}}, \href{http://dx.doi.org/10.1007/JHEP02(2015)116}{\emph{JHEP}
  {\bfseries 02} (2015) 116},
  [\href{https://arxiv.org/abs/1409.5656}{{\ttfamily 1409.5656}}].

\bibitem{Adamo:2017}
T.~Adamo, E.~Casali, L.~Mason and S.~Nekovar, \emph{{Amplitudes on curved
  backgrounds from ambitwistor strings}}, {\emph{In preparation} }.

\bibitem{Adamo:2014yya}
T.~Adamo, E.~Casali and D.~Skinner, \emph{{Perturbative gravity at null
  infinity}},
  \href{http://dx.doi.org/10.1088/0264-9381/31/22/225008}{\emph{Class. Quant.
  Grav.} {\bfseries 31} (2014) 225008},
  [\href{https://arxiv.org/abs/1405.5122}{{\ttfamily 1405.5122}}].

\bibitem{Geyer:2014lca}
Y.~Geyer, A.~E. Lipstein and L.~Mason, \emph{{Ambitwistor strings at null
  infinity and (subleading) soft limits}},
  \href{http://dx.doi.org/10.1088/0264-9381/32/5/055003}{\emph{Class. Quant.
  Grav.} {\bfseries 32} (2015) 055003},
  [\href{https://arxiv.org/abs/1406.1462}{{\ttfamily 1406.1462}}].

\bibitem{Adamo:2015fwa}
T.~Adamo and E.~Casali, \emph{{Perturbative gauge theory at null infinity}},
  \href{http://dx.doi.org/10.1103/PhysRevD.91.125022}{\emph{Phys. Rev.}
  {\bfseries D91} (2015) 125022},
  [\href{https://arxiv.org/abs/1504.02304}{{\ttfamily 1504.02304}}].

\bibitem{Aichelburg:1970dh}
P.~C. Aichelburg and R.~U. Sexl, \emph{{On the Gravitational field of a
  massless particle}}, \href{http://dx.doi.org/10.1007/BF00758149}{\emph{Gen.
  Rel. Grav.} {\bfseries 2} (1971) 303--312}.

\bibitem{Penrose:1972xrn}
R.~Penrose, \emph{{The geometry of impulsive gravitational waves}},  in
  \emph{General relativity: Papers in honour of J.L. Synge} (L.~O'Raifeartaigh,
  ed.), pp.~101--115.
\newblock 1972.

\bibitem{Dray:1984ha}
T.~Dray and G.~'t~Hooft, \emph{{The Gravitational Shock Wave of a Massless
  Particle}}, \href{http://dx.doi.org/10.1016/0550-3213(85)90525-5}{\emph{Nucl.
  Phys.} {\bfseries B253} (1985) 173--188}.

\bibitem{Klimcik:1988az}
C.~Klimcik, \emph{{Quantum Field Theory in Gravitational Shock Wave
  Background}},
  \href{http://dx.doi.org/10.1016/0370-2693(88)90632-6}{\emph{Phys. Lett.}
  {\bfseries B208} (1988) 373--380}.

\bibitem{Ferrari:1988cc}
V.~Ferrari, P.~Pendenza and G.~Veneziano, \emph{{Beamlike Gravitational Waves
  and Their Geodesics}}, \href{http://dx.doi.org/10.1007/BF00758938}{\emph{Gen.
  Rel. Grav.} {\bfseries 20} (1988) 1185--1191}.

\bibitem{Arefeva:1974jv}
I.~{\relax Ya}. Arefeva, L.~D. Faddeev and A.~A. Slavnov, \emph{{Generating
  Functional for the s Matrix in Gauge Theories}},
  \href{http://dx.doi.org/10.1007/BF01038094}{\emph{Theor. Math. Phys.}
  {\bfseries 21} (1975) 1165}.

\bibitem{Jevicki:1987ax}
A.~Jevicki and C.-k. Lee, \emph{{The S Matrix Generating Functional and
  Effective Action}},
  \href{http://dx.doi.org/10.1103/PhysRevD.37.1485}{\emph{Phys. Rev.}
  {\bfseries D37} (1988) 1485}.

\bibitem{Rosly:1996vr}
A.~A. Rosly and K.~G. Selivanov, \emph{{On amplitudes in selfdual sector of
  Yang-Mills theory}},
  \href{http://dx.doi.org/10.1016/S0370-2693(97)00268-2}{\emph{Phys. Lett.}
  {\bfseries B399} (1997) 135--140},
  [\href{https://arxiv.org/abs/hep-th/9611101}{{\ttfamily hep-th/9611101}}].

\end{thebibliography}\endgroup
\bibliographystyle{JHEP}

\end{document}